%% file: main_journal2.tex
\documentclass[journal=jacsat,manuscript=article]{achemso}

\usepackage[T1]{fontenc}
\usepackage{geometry}
\usepackage{setspace}

\mciteErrorOnUnknownfalse

\usepackage{indentfirst}

\usepackage{chemformula} % Formulas using \ch{}
\usepackage[version=3]{mhchem} % Formulas using \ce{}
\SectionNumbersOn
\input{math_commands.tex}

\usepackage{hyperref}
\usepackage{url}
\usepackage{graphicx}
\graphicspath{{figures/}} %%图片路径
\usepackage{algorithm}
\usepackage{algpseudocode}
\usepackage{booktabs} % 导入三线表需要的宏包
\usepackage{multirow}
\usepackage{graphicx}
\usepackage{wrapfig}

\author{Yan Wang}
\affiliation[Unknown University]
{School of Mathematical Sciences, Tongji University, Shanghai, China}
\altaffiliation{This work was done during a visit at Chalmers University of Technology, Gothenburg, Sweden.}
\author{Hao Wu}
\affiliation[Unknown University]
{School of Mathematical Sciences, Institute of Natural Sciences and MOE-LSC, Shanghai Jiao Tong University, Shanghai, China}
\email{hwu81@sjtu.edu.cn  (H. Wu)}
\author{Simon Olsson}
\affiliation[Unknown University]
{Department of Computer Science and Engineering, Chalmers University of Technology and University of Gothenburg, SE-41296 Gothenburg, Sweden}
\email{simonols@chalmers.se  (S. Olsson)}

\title{Marginal Girsanov Reweighting: Stable Variance Reduction for Long-Timescale Dynamics from Biased Simulation}

\abbreviations{IR,NMR,UV}

\begin{document}

% \maketitle

\begin{abstract}
Recovering unbiased kinetic and thermodynamic observables from the enhanced sampling simulations is a central challenge in rare-event sampling. Classical Girsanov Reweighting (GR) offers a principled solution by yielding exact pathwise probability ratios between biased and unbiased processes. However, the variance of GR weights grows rapidly with time, rendering it impractical for long-horizon reweighting. We introduce Marginal Girsanov Reweighting (MGR), which mitigates variance explosion by marginalizing over intermediate paths, producing stable and scalable weights for long-timescale dynamics. Experiments on various molecular dynamics systems demonstrate that MGR accurately recovers unbiased kinetic properties from trajectories generated under both umbrella sampling and metadynamics biases.

\end{abstract}

\section{Introduction}

Reweighting strategies play a central role in computational chemistry~\cite{noe2019boltzmann}. 
% mathematics~\cite{beskos2005exact}, finance~\cite{pascucci2011pde}, and machine learning~\cite{domke2018importance}. 
Especially in molecular dynamics (MD), direct sampling of the target distribution is often impractical because rare events and slow mixing make relevant transitions exceedingly unlikely to observe. Enhanced sampling methods such as metadynamics~\cite{barducci2008well}, umbrella sampling~\cite{torrie1977nonphysical}, and replica-exchange~\cite{sugita1999replica} address this by biasing the simulation to explore otherwise inaccessible regions of configuration space. Because these methods deliberately distort the target distribution, reweighting is not merely useful but essential to translate the biased statistics back into physically meaningful, unbiased estimates~\cite{kamenik2022enhanced}.

For thermodynamic properties, well-established methods such as the weighted histogram analysis method (WHAM)~\cite{gallicchio2005temperature,souaille2001extension} and the multistate Bennett acceptance ratio (MBAR)~\cite{shirts2008statistically} provide efficient estimators of equilibrium free energies from multiple simulation windows. 
Recent work has also employed machine learning to estimate energy surface across thermodynamic states~\cite{dibak2022temperature,wang2022data,moqvist2025thermodynamic, Invernizzi2022}. 
However, recovering kinetic properties from biased simulations is considerably more challenging. The standard framework for characterizing long-time dynamics is the Markov state model (MSM)~\cite{prinz2011markov,husic2018markov}, which requires unbiased transition counts between discrete states.
% When trajectories are generated under a biased potential, each observed transition must be reweighted to recover the unbiased transition matrix. 
A family of TRAM methods \cite{wu2014statistically,mey2014xtram,wu2016multiensemble,galama2023stochastic} formulate multiensemble Markov models that combine stationary free-energy estimation with transition count analysis to extract both thermodynamic and kinetic properties from simulations at multiple thermodynamic states.
The DHAM framework \cite{rosta2015free,stelzl2017dynamic} takes a related approach, reconstructing rate matrices from biased simulations to recover kinetics across thermodynamic states. 
However, these approaches reweight stationary distributions across ensembles but not the underlying transition probabilities \cite{klein2023timewarp,schreiner2023implicit,diez2026transferable}. As a consequence, they cannot reconstruct kinetic properties at a target thermodynamic state without simulation data collected at that state, and do not provide explicit per-frame reweighting factors for arbitrary dynamical observables. 

Maximum Caliber-based approaches reweight transition probabilities through constrained path entropy maximization and have been extended to collective variable spaces and non-equilibrium steady states~\cite{bause2019microscopic,bause2021reweighting}. A recent work learns the eigenfunctions of the transfer operator directly from biased trajectories~\cite{NEURIPS2024_89edef87}, but is tailored to eigenfunction recovery. However, neither approach provides explicit per-frame reweighting factors at the integrator level, limiting their applicability to the analysis of general dynamical observables. 

% Methods such as xTRAM~\cite{mey2014xtram} combine ideas from MBAR with transition count data to estimate both thermodynamic and kinetic quantities from replica-exchange simulations, but they rely on specific exchange protocols and global equilibrium assumptions.

Across these settings, a principled approach to computing importance weights over path space is provided by Girsanov's theorem~\cite{girsanov1960transforming}. Donati et al.~\cite{donati2017girsanov} first applied Girsanov reweighting (GR) to molecular dynamics, showing that pathwise likelihood ratios can be computed on-the-fly and combined with Boltzmann reweighting to construct Markov state models of a target potential from biased trajectories, including metadynamics simulations~\cite{donati2018girsanov}. The theoretical framework was subsequently extended to underdamped Langevin dynamics by deriving exact path probability ratios for practical MD integrators~\cite{kieninger2021path,kieninger2023girsanov,keller2026internal}. Then comprehensive reviews are given in Refs.~\cite{donati2022review,keller2024dynamical}. Beyond recovering unbiased kinetics, GR has been combined with data-driven discovery of slow collective variables for adaptive enhanced sampling~\cite{shmilovich2023girsanov}, applied to optimize force field parameters by imposing kinetic constraints~\cite{bolhuis2023optimizing}, generalized to construct MSMs from multiensemble biased simulations with time-dependent bias potentials~\cite{zhang2026pi}, and extended to first-principles molecular dynamics~\cite{jahnigen2026implementation}. A practical, open-source implementation of GR is available in OpenMM with a reweighted MSM estimator in Deeptime~\cite{schafer2024implementation}.

% Across these settings, a principle approach to computing importance weights over path space is provided by Girsanov's theorem~\cite{girsanov1960transforming}. Girsanov reweighting (GR) yields a pathwise likelihood ratio that enables a change of measure between SDEs sharing the same diffusion coefficient but differing in drift~\cite{donati2017girsanov,donati2022review, kieninger2021path,keller2024dynamical,kieninger2023girsanov}. Concretely, GR assigns each transition segment a weight derived from the log-ratio of trajectory densities under the unbiased and biased dynamics, thereby reweighting the short-time propagator---i.e., the transition operator---without requiring the generation of new trajectories. A practical, open-source implementation of GR has recently been made available as a plugin for the OpenMM molecular simulation toolkit~\cite{schafer2024implementation}, facilitating its integration into existing enhanced-sampling workflows.

Despite its elegance, a fundamental limitation of GR is that the variance of the pathwise weights grows exponentially with trajectory length and system dimensionality as introduced in Section~\ref{sec:gr}, making direct path reweighting impractical for long time scales and large systems. However, kinetic modeling of molecular systems typically operates at the level of transition probabilities, which are marginal quantities obtained by integrating over all intermediate paths between two endpoints. This marginalization can, in principle, tame the variance explosion by summing over all intermediate paths.

% To exploit this insight, we propose a machine learning–based approach: {\bf Marginal Girsanov Reweighting (MGR)}. Instead of relying on full-path Girsanov weights, which become numerically unstable for long trajectories, MGR learns marginal density ratios between end-points of trajectories. The key idea is to leverage accurate short-lag Girsanov weights and iteratively compose them into longer-lag ratios using neural classifiers. We formulate ratio estimation as a binary classification problem, where a neural network distinguishes between samples from reference and target distributions \cite{menon2016linking,choi2021featurized}. By combining the mathematical foundation of Girsanov reweighting with the flexibility of neural ratio estimation, MGR enables reliable inference across domains where traditional estimators break down.

To exploit this insight, we propose a machine learning–based approach: {\bf Marginal Girsanov Reweighting (MGR)}. Instead of using the long-lag full-path Girsanov weight directly, MGR learns the marginal density ratio by integrating out the intermediate path degrees of freedom. By the Rao–Blackwell theorem \cite{blackwell1947conditional}, this marginal ratio has provably equal or lower variance than its path-level counterpart, yet can be applied to reweight kinetic observables for long-timescale dynamics effectively. In practice, MGR uses the numerically stable short-lag Girsanov weights as training signal and iteratively composes them into longer-lag marginal ratios via neural classifiers trained on a binary classification objective \cite{menon2016linking,cranmer2020frontier}.
To summarize:

\begin{itemize}
    \item We propose MGR, an iterative learning approach, which estimates transition-based density ratios by marginalizing over intermediate paths.
    
    \item We formulate MGR as a binary classification task, yielding numerically stable estimates of the reweighting factors.
    
    \item We demonstrate the effectiveness of MGR by recovering unbiased thermodynamic and kinetic properties from biased MD simulations, e.g. umbrella sampling and metadynamics.
\end{itemize}

\begin{figure}[t!]
\begin{center}
\includegraphics[width=1.\textwidth]{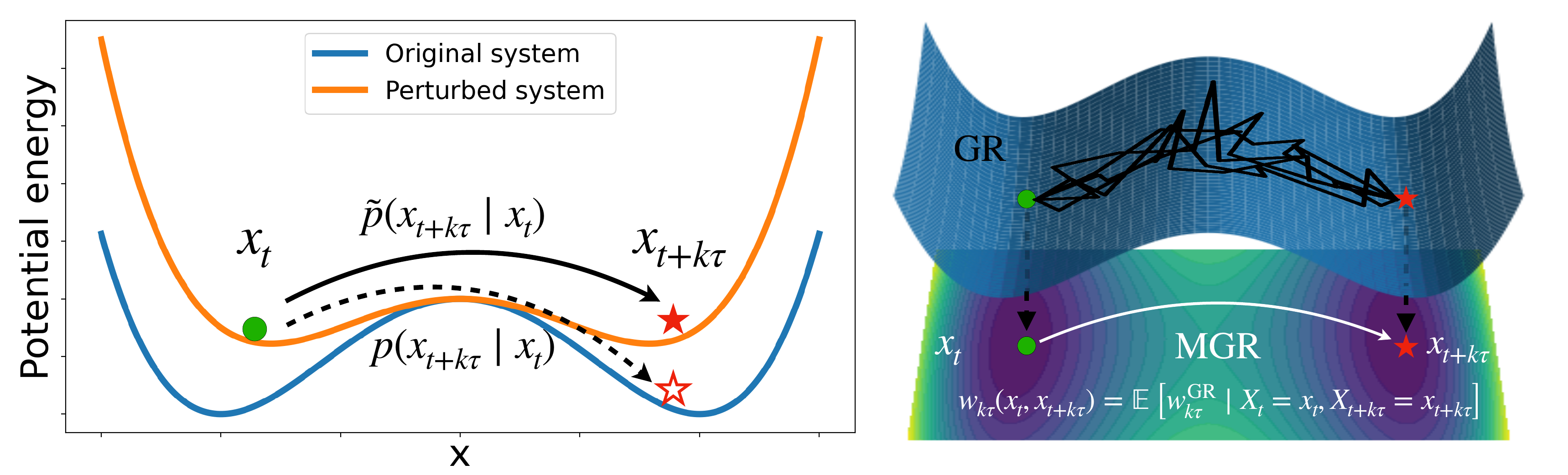}
\end{center}
\vspace{-4mm}
\caption{Marginal Girsanov Reweighting. For the given pairs, MGR defines the marginal weight as expectation of pathwise Girsanov reweighting factors as introduced in Eq.~\ref{eq:ratio-transition}.} \label{fig:MGR}

\end{figure}

\section{Theory}

\subsection{Brownian dynamics and mixing}

A common model for molecular dynamics is the overdamped Langevin equation:
% A diffusion process described by the overdamped Langevin equation satisfies
\begin{eqnarray}\label{eq:unbiasedsde}
    \mathrm d X_t = - \nabla V(X_t) \mathrm d t + \sigma \,\mathrm d W_t, \quad X_0 = x_0,
\end{eqnarray}
where $X_t \in \mathbb R^d$ denotes the state of the stochastic process at time $t$, $V(\cdot): \mathbb R^d \rightarrow  \mathbb R$ is the potential energy, $\sigma \in \mathbb R$ is the diffusion coefficient, and $W_t \in \mathbb R^d$ is a standard Wiener process. In a fixed time horizon $t\in [0, T]$, the associated path probability measure induced by Eq.~\ref{eq:unbiasedsde} is denoted by $\mu$.

However, direct simulation of the unbiased dynamics in Eq.~\ref{eq:unbiasedsde} often suffers from poor sampling efficiency, as high free-energy barriers separate metastable states and make transitions between them rare~\cite{vanden2010transition}. Enhanced sampling methods address this by adding a bias potential to the original force field, effectively lowering these barriers and accelerating transitions. The biased dynamics, e.g., umbrella sampling \cite{kastner2011umbrella} and metadynamics \cite{barducci2008well}, can be described as a system with a perturbed potential $\tilde V(\cdot, t) = V(\cdot) + U(\cdot, t)$, where $U(\cdot, t): \Gamma \rightarrow \mathbb{R}$ is the perturbation. The associated path probability measure is denoted by $\tilde \mu$. The extension to the underdamped Langevin is introduced in Appendix~\ref{appendix:underdamped}.

Once the biased simulation has achieved sufficient mixing, equilibrium properties can be recovered through standard reweighting of the stationary distribution. Beyond thermodynamic reweighting \cite{mey2014xtram,wu2016multiensemble}, Girsanov's theorem provides a route to reweight kinetic properties in path space by evaluating the Radon--Nikodym derivative between two path measures, thereby recovering dynamical observables from the biased simulation. Before detailing this procedure, we first clarify the target quantity that the reweighting should deliver.

\subsection{Kinetic Reweighting}\label{sec:setting}
In enhanced sampling simulations, the full trajectory contains every integration step along the biased dynamics. However, to characterize molecular kinetics \cite{prinz2011markov,tiwary2015kinetics,mardt2018vampnets}—such as folding rates, ligand binding timescales, or conformational transition probabilities—one typically only needs to know how likely the system is to move from one configuration $x_t$ to another $x_{t+\tau}$ over a lag time $\tau$ ~\cite{klein2023timewarp,schreiner2023implicit,diez2026transferable}. The central object is therefore the transition probability.

Let $p\left(x_{t+\tau} \mid x_t \right)$ denote the transition probability induced by the unbiased process in Eq.~\ref{eq:unbiasedsde}, and let $\tilde{p}\left(x_{t+\tau} \mid x_t\right)$ denote the transition probability under the enhanced sampling dynamics. To recover unbiased kinetics from biased simulations, we seek the reweighting factor $w_{\tau}(x_t, x_{t+\tau}) = \frac{p\left(x_{t+\tau} \mid x_t \right)}{\tilde p\left(x_{t+\tau} \mid x_t\right)}$.

Computing this ratio directly requires knowledge of both conditional densities, which are generally intractable. However, the ratio simplifies when expressed in terms of joint distributions. Let $\rho(x_t)$ denote the marginal density at time $t$, and let $\rho_{\tau}(x_t, x_{t+\tau})$ and $\tilde{\rho}_{\tau}(x_t, x_{t+\tau})$ denote the joint densities under the unbiased and biased processes respectively. The reweighting factor then becomes
\begin{eqnarray*}
    w_{\tau}(x_t, x_{t+\tau}) = \frac{p \left(x_{t+\tau} \mid x_t\right) \rho \left(x_t \right)}{\tilde p\left(x_{t+\tau} \mid x_t\right)\rho \left(x_t\right)} = \frac{\rho_{\tau} \left(x_t, x_{t+\tau}\right)}{\tilde \rho_{\tau} \left(x_t, x_{t+\tau}\right)}.
\end{eqnarray*}
This representation highlights that the ratio of transition density ratio can be seen as the ratio of joint distributions. Furthermore, it admits a natural transformation in terms of the Radon–Nikodym derivative over path space
\begin{eqnarray}
    w_{\tau}(x_t, x_{t+\tau}) = \mathbb E_{\tilde \mu}\left[ \frac{\mathrm d \mu}{\mathrm d \tilde \mu}(\mathbf x_{t, \tau}) \mid X_t=x_t, X_{t+\tau}=x_{t+\tau} \right]. \label{eq:ratio-transition}
\end{eqnarray}
Here, $\mathbf x_{t, \tau} = \{x_s \}_{s=t}^{t+\tau}$ denotes the trajectory path from $x_t$ to $x_{t+\tau}$, $t\in[0, T-\tau]$. Detailed proof can be found in Appendix~\ref{appendix:transition}.
Notably, Eq.~\ref{eq:ratio-transition} takes the form of a conditional expectation of the pathwise likelihood ratio. By the Rao–Blackwell theorem \cite{blackwell1947conditional}, this conditional expectation has variance no greater than that of the full-path weight $\frac{\mathrm d \mu}{\mathrm d \tilde \mu}(\mathbf x_{t, \tau})$ itself, 
\begin{eqnarray*}
\mathrm{Var}\!\left[w_{\tau}(x_t, x_{t+\tau})\right] \leq \mathrm{Var}\!\left[\frac{\mathrm d \mu}{\mathrm d \tilde \mu}(\mathbf x_{t,\tau})\right].
\end{eqnarray*}
This variance reduction is the central motivation for working with marginal density ratios rather than path-level weights.

Physically, this reweighting factor is used to construct unbiased Markov state models by reweighting the cross-correlation matrix estimated from biased trajectories (Eq.~\ref{eq:msm_count}). A classical tool for evaluating this likelihood ratio is Girsanov Reweighting, which we introduce next.

\subsection{Girsanov Reweighting Theory} \label{sec:gr}
A common approach to compute the Radon–Nikodym derivative of $\mu$ respect to $\tilde\mu$ is provided by Girsanov’s theorem \cite{girsanov1960transforming,donati2017girsanov}. If two diffusion processes share the same diffusion coefficient but have different drifts, their path measures can be transformed from one to the other.

Here, we consider the trajectory segment under the enhanced sampling dynamics $\mathbf x_{t, \tau} = \{x_s \}_{s=t}^{t+\tau}$. Using an Euler–Maruyama discretization, the corresponding discrete-time trajectory $\left\{x_t=x^0, x^1, x^2, \ldots, x_{t+\tau}=x^N\right\}$ is observed with discretization step $\Delta t=\tau/N$. Then the likelihood ratio between the unbiased path $\mu$ and the enhanced sampling path $\tilde \mu$, conditional on the same starting state $x_t$, can be calculated as

\begin{align}
\log w_{\tau}^{\text{GR}}(\mathbf x_{t, \tau}) &= \log \frac{\mathrm{d} \mu}{\mathrm{~d} \tilde{\mu}}(\mathbf x_{t, \tau}|x_t) \nonumber \\
&\approx \sum_{k=0}^{N-1}\left(\frac{\nabla U(x^k, t) ^{\top}}{\sigma} \sqrt{\Delta t} \xi^k-\frac{\Delta t}{2}\left\|\frac{\nabla U(x^k, t)}{\sigma}\right\|^2\right), \label{eq:gr-discrete}
\end{align}

where $\sqrt{\Delta t}\,\xi^k = \dfrac{x^{k+1}-x^k+\nabla\tilde{V}\!\left(x^k, t\right)\Delta t}{\sigma}\stackrel{\text { i.i.d. }}{\sim} \mathcal{N}\left(0, \Delta t\cdot I_d\right)$
represents the Wiener increment associated with the simulation step $x^k \rightarrow x^{k+1}$ under the perturbed dynamics. The derivation above assumes overdamped Langevin dynamics for notational simplicity. The extension to underdamped Langevin dynamics, where both position $x_t$ and momentum $v_t$ are propagated, is detailed in Appendix~\ref{appendix:underdamped}.

Eq.~\ref{eq:gr-discrete} provides a solution for recovering original properties from perturbed simulations. However, a fundamental difficulty arises in practice: computing $\log w_{\tau}^{\text{GR}}$ over long time horizons $\tau$ introduces an accumulation of noise terms, e.g. $\xi^k$ in Eq.~\ref{eq:gr-discrete}, which causes the variance to grow with trajectory length. Upon exponentiation, the weights will explode or vanish. Detailed analysis can be found in Appendix~\ref{appendix:gr-variance}. 
Yet kinetic modeling of molecular systems typically operates at the level of transition probabilities, which are marginal quantities obtained by integrating over all intermediate paths between two endpoints (see Section~\ref{sec:setting} for details). Unlike pathwise Girsanov weights, these marginal ratios sum over all intermediate paths, which can in principle tame the variance explosion. Motivated by this observation, we develop a model that directly estimates marginal reweighting factors for transition probabilities, yielding stable and accurate weights even for long time intervals and large systems.

\section{Method} \label{sec:MGR}
As established in Section~\ref{sec:setting}, the kinetic reweighting factor $w_{\tau}(x_t, x_{t+\tau})$ is a marginal quantity that integrates over all intermediate paths, and is therefore in principle more stable than the pathwise Girsanov weight. However, this marginal ratio involves intractable conditional densities and cannot be computed in closed form. We propose \textbf{Marginal Girsanov Reweighting (MGR)}, a machine learning approach that bypasses the explicit density computation by learning the marginal reweighting factor directly from simulation data. The key idea is to use the short-lag Girsanov weights—where the variance remains controlled—as supervised signals, and to extend the estimates to longer lag times through a self-consistent training procedure. We describe the algorithm below.

\subsection{Training Algorithm} \label{sec:algorithm}
We denote by $w_{k \tau}(x_t, x_{t+k\tau})$ the likelihood ratio from configuration $x_t$ at time $t$ to $x_{t+k\tau}$ at time $t+k\tau$. For the case $k=1$, i.e., over a short lag time, Girsanov reweighting $w_{\tau}^{\text{GR}}$ in Eq.~\ref{eq:gr-discrete} provides a relatively stable and pathwise estimate of the ratio. Our goal is to develop a Marginal Girsanov Reweighting approach that can reliably estimate the ratio for longer lag times with $k \gg 1$.

Our method, MGR, adopts an iterative training strategy based on either a long discretized simulation trajectory or multiple discretized trajectories under perturbed path $\tilde \mu$. Suppose the ratio $w_{(k-1)\tau}$ has already been obtained. Then the ratio $w_{k\tau}$ at lag time $k\tau$ can be constructed as follows:

\begin{figure}[t!]
\begin{center}
\includegraphics[width=0.5\textwidth]{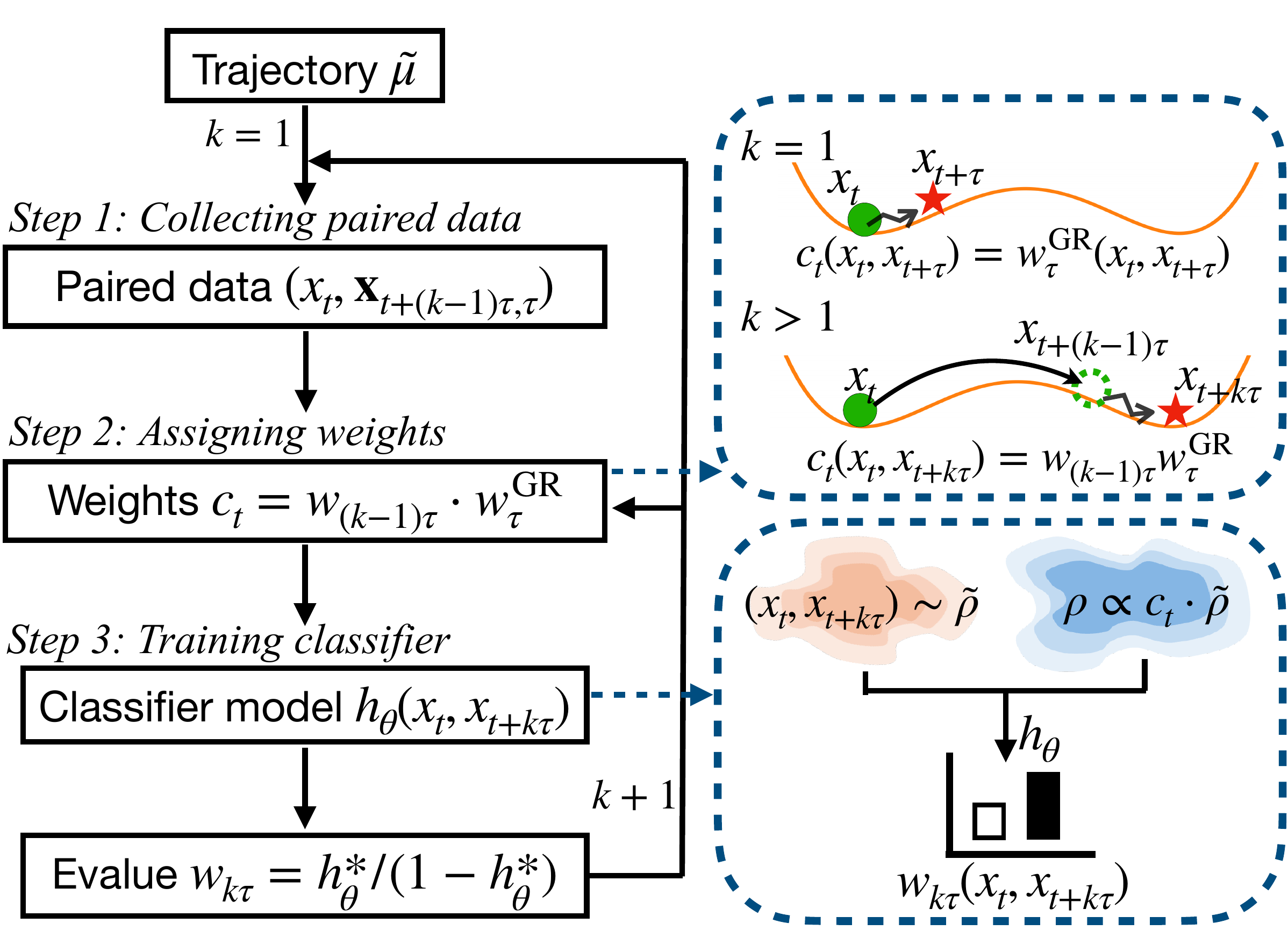}
\end{center}
\vspace{-3mm}
\caption{The training algorithm of MGR as illustrated in Sections~\ref{sec:MGR}.} \label{fig:MGR-alg}
\end{figure}

%\textbf{\textit{Step 1.}} We collect all paired data $\left\{\left(x_t, x_{t+k \tau}\right)\right\}_{t=0}^{T-k\tau}$ from the perturbed simulation trajectory.
\textbf{\textit{Step 1.}} We collect pairs of the form 
$\bigl(x_t,\, \mathbf{x}_{t+(k-1)\tau,\tau}\bigr)$ for $t \in [0,\, T-k\tau]$ 
from the perturbed simulation trajectory, 
where each pair consists of a state and a trajectory segment.

\textbf{\textit{Step 2.}} In this step, we construct approximations of perturbed and original distributions of $(x_t,x_{t+k\tau})$ based on the following identity:
\begin{align*}
\mathbb{E}_{\rho_{k\tau}(x_{t},x_{t+k\tau})}\!\left[O(x_{t},x_{t+k\tau})\right] 
 = & \mathbb{E}_{\rho_{(k-1)\tau}(x_{t},x_{t+(k-1)\tau}) \cdot \mu(\mathbf{x}_{t+(k-1)\tau,\tau}|x_{t+(k-1)\tau})}
    \!\left[O(x_{t},x_{t+k\tau})\right]\\
 = & \mathbb{E}_{\tilde{\rho}_{(k-1)\tau}(x_{t},x_{t+(k-1)\tau}) \cdot \tilde{\mu}(\mathbf{x}_{t+(k-1)\tau,\tau}|x_{t+(k-1)\tau})}
    \!\left[c_{t}\, O(x_{t},x_{t+k\tau})\right],
\end{align*}
for any bounded measurable test function 
$O(\cdot, \cdot): \mathbb{R}^d \times \mathbb{R}^d \rightarrow \mathbb{R}$. The first equality holds because $\rho_{k\tau}$ can be regarded as the marginal distribution of $(x_{t},x_{t+k\tau})$ 
defined by $\rho_{(k-1)\tau}(x_{t},x_{t+(k-1)\tau}) \cdot \mu(\mathbf{x}_{t+(k-1)\tau,\tau}\mid x_{t+(k-1)\tau})$, 
and the second equality follows from the principle of importance sampling, with weight
\begin{equation}\label{eq:c_t}
c_t = w_{(k-1)\tau}(x_t, x_{t+(k-1)\tau}) \cdot
w_\tau^{\text{GR}}\!\left(\mathbf{x}_{t+(k-1)\tau,\tau}\right).
\end{equation}
Since $w_{(k-1)\tau}$ is inherited from the previous iteration 
and $w_\tau^{\text{GR}}$ can be computed as described in Section~\ref{sec:gr}, 
the pairs collected in \textit{Step~1} can be used to approximate the perturbed and original joint distributions 
$\tilde\rho_{k\tau}$ and $\rho_{k\tau}$ as

\begin{eqnarray*}
\tilde \rho_{k\tau}(x, y) &\approx& \frac{1}{T-k\tau}\sum_t \delta(x-x_t)\, \delta(y-x_{t+k\tau}), \\
\rho_{k\tau}(x, y) &\approx& \sum_t \frac{c_t \, \delta(x-x_t)\, \delta(y-x_{t+k\tau})}{\sum_{t'} c_{t'}},
\end{eqnarray*}

where $\delta$ denotes the Dirac delta function.
When accurate estimates of $w_{(k-1)\tau}$ are available and the dataset size is sufficiently large, 
the above approximations can be shown to be consistent. 
A detailed proof is provided in Appendix~\ref{appendix:iterativetraining}.

\textbf{\textit{Step 3.}} Finally, using the joint distributions $\tilde \rho_{k\tau}(x, y)$, $\rho_{k\tau}(x, y)$ in \textit{Step 2}, we approximate the marginal weight $w_{k \tau}(x, y)$ by a classifier-based density ratio estimator (see Section~\ref{sec:ratio} for details). The inferred ratio from the optimal binary classifier is then used for the next iteration.

This three-step procedure effectively extends short-time pathwise Girsanov weights to long-time marginal ratio. First, we train the model $w_{\tau}$ using the short $\tau$-lag Girsanov weights. Then, we iterate three steps above to progressively learn $w_{k\tau}$ until $k\gg 1$. The workflow is summarized in Figure~\ref{fig:MGR-alg}. In the next subsection, we introduce the ratio estimation method used in MGR \textit{Step 3}.

%This iterative strategy trains a model that depends on endpoint pairs $\{(x_t, x_{t+k\tau})\}$. Thus, the model can implicitly learn the path integrals and absorbs the variance from intermediate noise, which ensures scalability to long lag times and high-dimensional systems. 

\subsection{Classifier-based Density Ratio Estimation} \label{sec:ratio}
Many machine learning methods have been proposed for density ratio estimation \cite{menon2016linking, choi2021featurized, choi2022density, yu2025density}. A widely used approach is probabilistic classification \cite{menon2016linking}, which reformulates likelihood ratio estimation as a binary classification task. In this setting, a binary classifier with a sigmoid output $h(\cdot, \cdot): \mathbb R^d \times \mathbb R^d \rightarrow [0, 1] $ is trained to discriminate between paired samples $(x_t, x_{t+k\tau})$ drawn from the perturbed distribution $\tilde \rho_{k\tau}$ and the original distribution $\rho_{k\tau}$. 

We define the optimal classifier $h_{\theta}^*$ as the probability that a given pair $(x_t, x_{t+k\tau})$ comes from $\rho_{k\tau}$, i.e., $h_{\theta}^*(x_t, x_{t+k\tau})={\rho_{k \tau}(x_t, x_{t+k\tau})} / \left(\rho_{k \tau}(x_t, x_{t+k\tau})+\tilde \rho_{k \tau}(x_t, x_{t+k\tau})\right)$.
Then, the density ratio can be estimated as
\begin{eqnarray}
w_{k \tau}(x_t, x_{t+k\tau})=\frac{\rho_{k \tau}(x_t, x_{t+k\tau})}{\tilde \rho_{k \tau}(x_t, x_{t+k\tau})}=\frac{h_\theta^*(x_t, x_{t+k\tau})}{1-h_\theta^*(x_t, x_{t+k\tau})} . \label{eq:w_ktau}
\end{eqnarray}

Unlike standard density ratio estimation, where samples from both distributions are available and the ratio can be learned via cross-entropy loss, our setting in MGR is different. We only have samples from the perturbed distribution $\tilde \rho_{k\tau}$ together with the corresponding weight $c_t$ assigned to each sample. To address this, we employ a weighted cross-entropy loss:
\begin{eqnarray}
\mathcal{L}(\theta) = - \mathbb E_{t} \left[ c_t \log h_{\theta}(x_t, x_{t+k\tau}) + \log \left(1- h_{\theta}(x_t, x_{t+k\tau})\right) \right], \label{eq:weightedBCE}
\end{eqnarray}
where $(x_t, x_{t+k\tau})$ denotes the paired data collected from the perturbed trajectory with lagtime $k\tau$. The weights $c_t$ are constructed from the output of the previous model $w_{(k-1)\tau}$ and short-lag Girsanov reweighting $w_{\tau}^{\text{GR}}$ according to Eq.~\ref{eq:c_t}, and are normalized over the entire dataset. 

After sufficient training on either a single long trajectory or multiple trajectories, the estimator $w_{k\tau}$ in Eq.~\ref{eq:w_ktau} converges to the marginal ratio, and then serves as a component for the next iteration. Through the iterative training scheme, MGR can attain stable marginal ratios under long lag times and in complex systems. The complete training procedure is summarized in Algorithm~\ref{alg:MGR} and illustrated in Figure \ref{fig:MGR-alg}. For completeness, we compare a range of ablations and discuss other potential model choices (see Appendix~\ref{appendix:ratio}), but defer a more systematic investigation to future work.

% The complete training procedure is summarized in Algorithm x and illustrated in Figure \ref{fig:MGR}. While Girsanov reweighting yields a ratio along a specific simulation path, the MGR model $h_{\theta}$ receives only the endpoint pairs $(x_t, x_{t+k\tau})$ as input and is independent of the intermediate path. Therefore, what the model effectively learns is the integral over all transition paths connecting the endpoints in Eq.~\ref{eq:ratio-transition}. Through the iterative training scheme, MGR is able to compute marginal ratios under long lag times and in complex systems.
\section{Experiments}

MGR is particularly useful for analyzing kinetic properties from biased MD \cite{donati2017girsanov,donati2022review}. In particular, it can be employed in the construction of MSMs \cite{prinz2011markov} at a given lag time.

In an MSM framework, the dynamics are characterized by a transition probability matrix $P_\tau$, whose entries are obtained by normalizing the cross-correlation matrix $C_{ij}(\tau)$. Under the unbiased measure $\mu$, this quantity is defined as $C_{ij}(\tau) = \mathbb{E}_{\mu}\left[\mathbf{1}_{B_i}(x_t)\,\mathbf{1}_{B_j}(x_{t+\tau})\right]$.
When trajectories are generated under a biased measure $\tilde{\mu}$, the unbiased cross-correlation can be recovered via importance reweighting:
\begin{equation}\label{eq:msm_count}
    C_{ij}(\tau) = \mathbb{E}_{\tilde{\mu}}\left[w_\tau(x_t, x_{t+\tau})\,
    \mathbf{1}_{B_i}(x_t)\,\mathbf{1}_{B_j}(x_{t+\tau})\right]
    \approx \frac{\sum_t w_\tau(x_t, x_{t+\tau})\,\mathbf{1}_{B_i}(x_t)\,
    \mathbf{1}_{B_j}(x_{t+\tau})}{\sum_t w_\tau(x_t, x_{t+\tau})},
\end{equation}
where $w_\tau(x_t, x_{t+\tau})$ denotes the marginal transition weight estimated by MGR. In the GR baseline~\cite{donati2017girsanov}, this weight is instead computed via pathwise Girsanov reweighting. Once the reweighted transition matrix is constructed, standard MSM analysis can be performe. Detailed information is provided in Appendix~\ref{appendix:experiments}. 
% In our experiments on alanine dipeptide and deca-alanine, MSMs and the MGR model are constructed on a set of pre-selected reaction coordinates rather than the full atomic configuration. This projection can be justified under standard assumptions on the reaction coordinates (see Appendix~\ref{appendix:rc-input} for details), and a systematic study of its effect on the reweighting accuracy is left for future work.

We primarily compare MGR with GR throughout the experiments. All numerical experiments in this work use a single biased system, for which GR is the only existing method that provides explicit per-transition reweighting factors without requiring global equilibrium. The extension of MGR to multi-ensemble settings, where multiple biased simulations are combined, is studied in a separate line of work~\cite{zhang2026pi}. 
Guidelines for selecting the training lag time and network architecture are provided in Appendices~\ref{appendix:lag-ablation} and~\ref{appendix:neural-ablation}. Computational cost is analyzed in Appendix~\ref{appendix:cost}.

We primarily adopt the following metrics to evaluate the quality of the recovered MSM.

\paragraph{Effective Sample Size (ESS).}
ESS quantifies the variance of the importance weights and serves as an indicator of reweighting stability~\cite{freeman1966kish}. Given weights $\{w_\tau(x_t, x_{t+\tau})\}_{t=1}^{M}$, the relative ESS is defined as
\begin{equation*}
    \mathrm{rESS}_\tau = \frac{\left(\sum_t w_\tau(x_t, x_{t+\tau})\right)^2}
    {M \sum_t w_\tau(x_t, x_{t+\tau})^2},
\end{equation*}
which takes values in $(0, 1]$. A higher value indicates reduced variance and more reliable statistical estimates.

\paragraph{Implied Timescales (ITS).}
Let $1 = \lambda_1(\tau) > \lambda_2(\tau) \geq \lambda_3(\tau) \geq \cdots$ be the leading 
eigenvalues of $P_\tau$. Each eigenvalue defines an implied timescale
\begin{equation*}
    t_i(\tau) = -\frac{\tau}{\ln \lambda_i(\tau)},
\end{equation*}
which measures the relaxation time of the $i$-th slow dynamical mode. Beyond inspecting individual timescales, we also track the aggregate relaxation measure
\begin{equation*}
    S_m(\tau) := \sum_{i=2}^{m+1} \lambda_i(\tau) 
    = \sum_{i=2}^{m+1} \exp\!\left(-\frac{\tau}{t_i(\tau)}\right).
\end{equation*}
For a well-estimated MSM, $S_m(\tau)$ decays approximately exponentially with $\tau$, reflecting the intrinsic relaxation spectrum of the dynamics. In practice, we report both dominant individual ITS and $S_m(\tau)$ to assess mode-wise and overall accuracy.

\paragraph{Dominant Eigenfunctions and Stationary Distribution.}
From the estimated MSM, one can extract the stationary distribution $\pi$ and the dominant eigenfunctions $\{\boldsymbol{\phi}_2, \boldsymbol{\phi}_3, \ldots\}$~\cite{prinz2011markov,scherer2015pyemma,hoffmann2022deeptime}. The former describes the long-time equilibrium and validates the thermodynamic properties. The latter encode the slow dynamical modes and validate the kinetics.

\subsection{One Dimensional Four Well}

\begin{figure}[b!]
\begin{center}
\includegraphics[width=1.\textwidth]{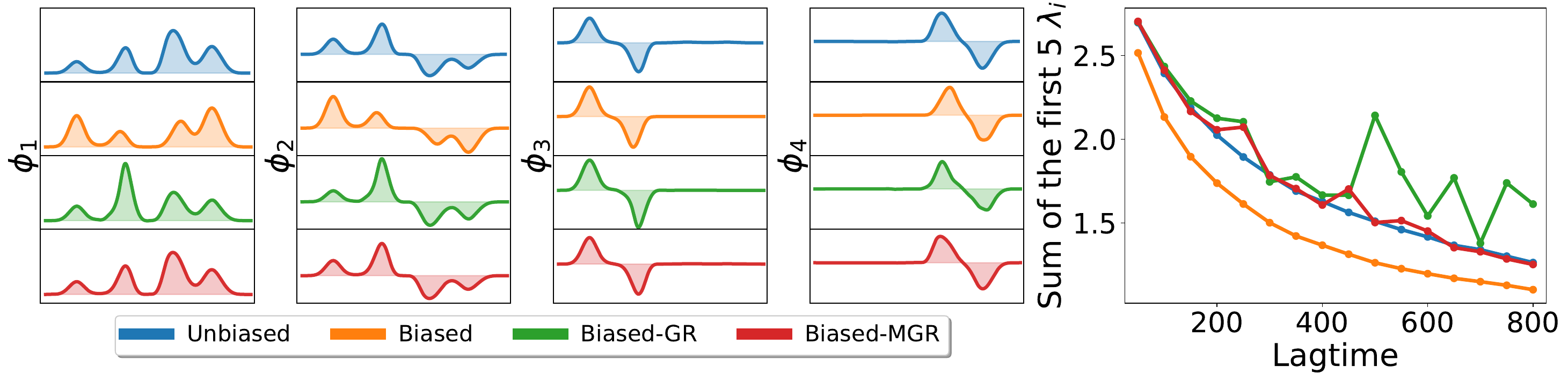}
\end{center}
\vspace{-4mm}
\caption{Results for the one-dimensional four-well potential. \textbf{Left:} Dominant left 
eigenfunctions ($\{\boldsymbol{\phi}_i\}$) of the transition matrix at lag time $300\Delta t$ 
($\boldsymbol{\phi}_1$ denotes the stationary distribution). Unbiased results serve as 
reference; biased, GR-reweighted, and MGR-reweighted results are compared. \textbf{Right:} Sum 
of the first five eigenvalues as a function of lag time, reflecting the intrinsic relaxation 
behavior of the dynamics.}
\label{fig:four-well}
\end{figure}

We first consider a one-dimensional four-well potential system \cite{prinz2011markov}, which serves as a prototypical example for testing reweighting methods. The unbiased energy landscape contains four metastable states separated by barriers, with the two intermediate wells located at higher energies. To accelerate sampling, we introduce a biased potential that lowers the energy of the two intermediate wells. Detailed energy function and simulation information can be found in Appendix \ref{appendix:fourwell}.

In this landscape, the first three slow kinetic modes are physically meaningful. The slowest mode (ITS component 2) corresponds to the global left--right rearrangement across the highest barrier. The next two modes capture exchanges within the left pair and right pair of wells, respectively. The remaining modes reflect fast relaxation and are not expected to be stable across lag times.

\begin{figure}[t!]
\begin{center}
\includegraphics[width=1\textwidth]{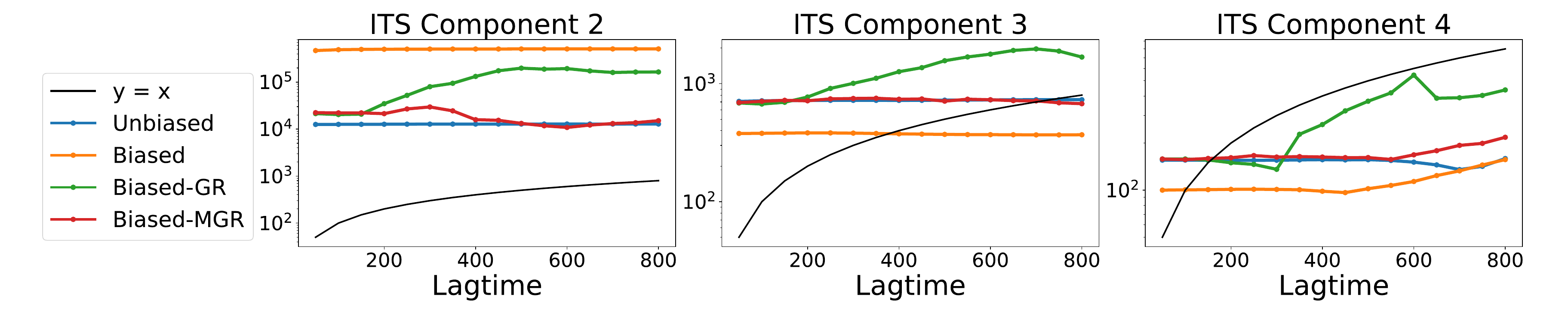}
\end{center}
\vspace{-6mm}
\caption{First three ITS as a function of lag time. Each panel reports $t_i(\tau)$ for mode $i = 2, \ldots, 4$ (log scale). MGR tracks the unbiased ITS with less fluctuation, while GR deviates at long lag times.}\label{fig:four-well-ITS}
\end{figure}

Figure~\ref{fig:four-well} (left) shows the dominant left eigenfunctions at lag time $300\Delta t$ ($\Delta t = 0.001$). GR provides partial correction but still exhibits clear deviations, particularly in the central wells, whereas MGR yields eigenfunctions in close agreement with the unbiased reference. The right panel compares the sum of the first five eigenvalues as a function of lag time. MGR closely follows the unbiased reference with smooth exponential decay, while GR suffers from strong fluctuations, indicating that MGR more accurately captures the intrinsic relaxation spectrum.

\begin{figure}[t!]
\begin{center}
\includegraphics[width=1\textwidth]{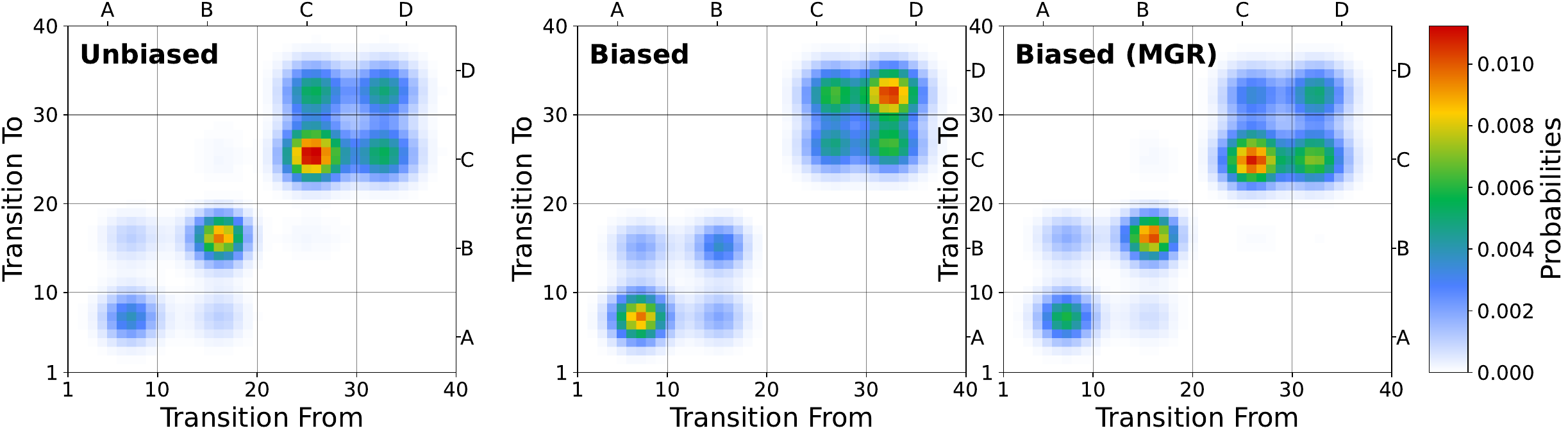}
\end{center}
\vspace{-4mm}
\caption{Density plot of the transfer operator for the four-well system on the discrete spatial range $[0, 40]$ at lag time $300\Delta t$.}
\label{fig:four-well-transition}
\end{figure}

Figure~\ref{fig:four-well1} (left) further reports the dominant right eigenfunctions, confirming consistent recovery by MGR. The right panel shows the relative ESS as a function of lag time: ESS for GR decays rapidly at long lag times, reflecting weight instability, whereas MGR maintains substantially higher ESS throughout. Figure~\ref{fig:four-well-ITS} reports the first three individual implied timescales. MGR tracks the unbiased ITS with markedly less fluctuation, while GR deviates at long lag times.

Finally, MGR can also be used to recover the transfer operator density, defined as
\begin{equation}
    \rho_{k\tau}(x, y) = \rho(x)\, p_{k\tau}(x, y) 
    = e^{-V(x)}\, w_{k\tau}(x, y)\, \tilde{\rho}_{k\tau}(x, y),
\end{equation}
where $\tilde{\rho}_{k\tau}$ denotes the biased joint density. Figure~\ref{fig:four-well-transition} displays the resulting density plots of the transfer operator, demonstrating that MGR successfully recovers the unbiased transition structure.

\subsection{Müller-Brown Potential}

We next consider the two-dimensional Müller--Brown potential, a classical benchmark for enhanced sampling methods \cite{laio2002escaping}. It has two principal minima, one of which is further divided into two sub-basins. The system thus exhibits two slow processes: the dominant one corresponding to inter-basin transitions and the second to intra-basin rearrangements. At low temperature, spontaneous barrier crossings are rare, motivating the use of a bias potential to accelerate sampling.

We simulate the overdamped Langevin dynamics at temperature $T=1$ (arbitrary units) with a time step of $\Delta t = 10^{-3}$. The bias is constructed adaptively via metadynamics~\cite{laio2002escaping}. Specifically, the bias potential takes the form
\begin{equation*}
    U(\mathbf{x}, t) = h \sum_{i=1}^{N_t} 
    \exp\!\left(-\frac{\|\mathbf{x} - \mathbf{x}_i\|^2}{2\sigma^2}\right),
\end{equation*}
where Gaussian kernels of height $h$ and width $\sigma$ are deposited at the current position every 500 time steps. After $3 \times 10^5$ steps, no further kernels are added, yielding a static bias for the remainder of the simulation.

We emphasize that MGR does \textbf{not} require the simulation to have reached equilibrium. We analyze this system under two regimes: an equilibrium simulation of $10^7$ steps, where the trajectory fully relaxes under the static bias, and a non-equilibrium simulation of only $10^6$ steps, where equilibration is incomplete.

\begin{figure}[t!]
\begin{center}
\includegraphics[width=1\textwidth]{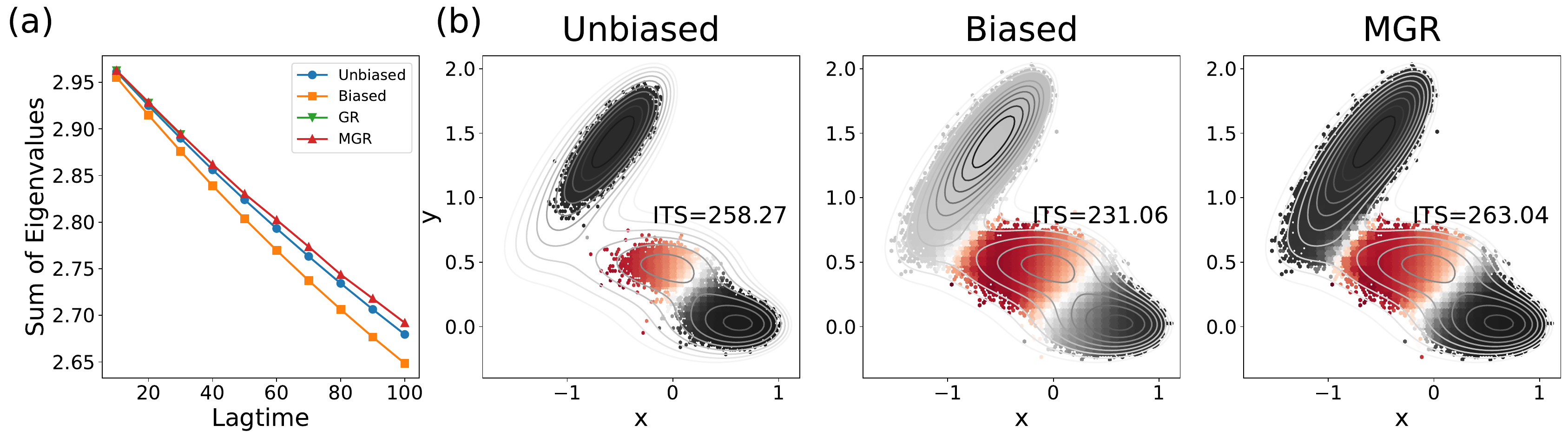}
\end{center}
\vspace{-4mm}
\caption{Müller-Brown potential: equilibrium regime ($10^7$ steps). \textbf{a)} Sum of the first three eigenvalues as a function of lag time. \textbf{b)} The second dominant right eigenfunctions at lag time $50\Delta t$ with corresponding ITS. Unbiased, biased, and MGR-reweighted results are compared.}
\label{fig:mb-eq}
\end{figure}
% \begin{figure}[t!]
% \begin{center}
% \includegraphics[width=1\textwidth]{figures/MB_short_eigenvectors.pdf}
% \end{center}
% \vspace{-4mm}
% \caption{Müller-Brown potential: non-equilibrium regime ($10^6$ steps). Layout identical to Figure~\ref{fig:mb-eq}. Despite incomplete equilibration, MGR recovers the unbiased eigenvalue 
% spectrum and eigenfunctions.}
% \label{fig:mb-neq}
% \end{figure}

Figure~\ref{fig:mb-eq} presents the results obtained from equilibrium biased simulations. Panel~(a) reports the sum of the first three eigenvalues as a function of lag time. MGR closely tracks the unbiased reference with smooth exponential decay, while the raw biased estimates deviate substantially. Panel~(b) displays the second dominant right eigenfunction at lag time $50\Delta t$ together with the corresponding implied timescale. This mode resolves the intra-basin rearrangement, and MGR accurately recovers both the eigenfunction shape and the implied timescale. We note that GR fails to construct a reversible MSM for lag times exceeding $30\Delta t$ due to excessive weight variance and is therefore omitted from the comparison. Results for the non-equilibrium regime are provided in Figure~\ref{fig:mb-neq}.

\begin{figure}[t!]
\begin{center}
\includegraphics[width=1\textwidth]{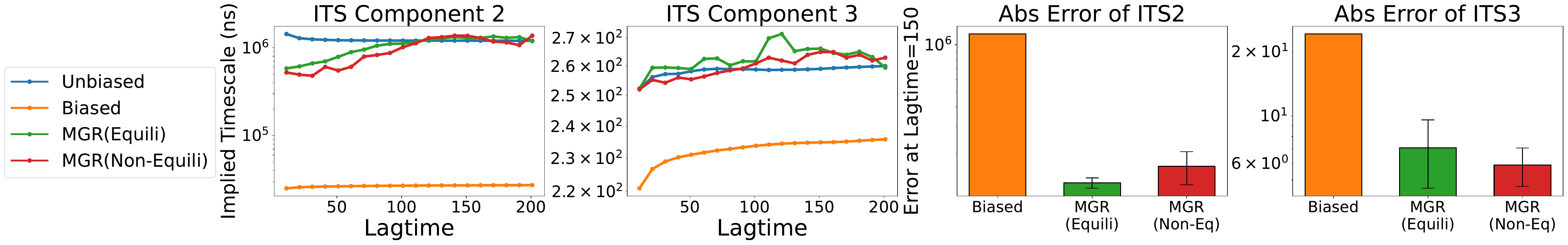}
\end{center}
\vspace{-4mm}
\caption{\textbf{Left:} First two ITS $t_i(\tau)$ for modes $i = 2, 3$ as a function of lag time under the equilibrium ($10^7$ steps) and non-equilibrium ($10^6$ steps) regimes of the Müller-Brown potential. \textbf{Right:} Mean and standard deviation of the ITS error at lagtime=150, evaluated against the unbiased reference values. Statistics were calculated over 5 independent replicates.}
\label{fig:mb-its}
\end{figure}

Figure~\ref{fig:mb-its} further reports the first two individual ITS as a function of lag time for both regimes. The unbiased reference is taken from the equilibrium simulation in both cases to provide a consistent baseline. In the both settings, MGR recovers the dominant two timescales accurately. This confirms that MGR preserves mode-wise kinetic accuracy regardless of whether the biased trajectory has equilibrated.

\subsection{Alanine Dipeptide}

\begin{figure}[t!]

\begin{center}
\includegraphics[width=1.\textwidth]{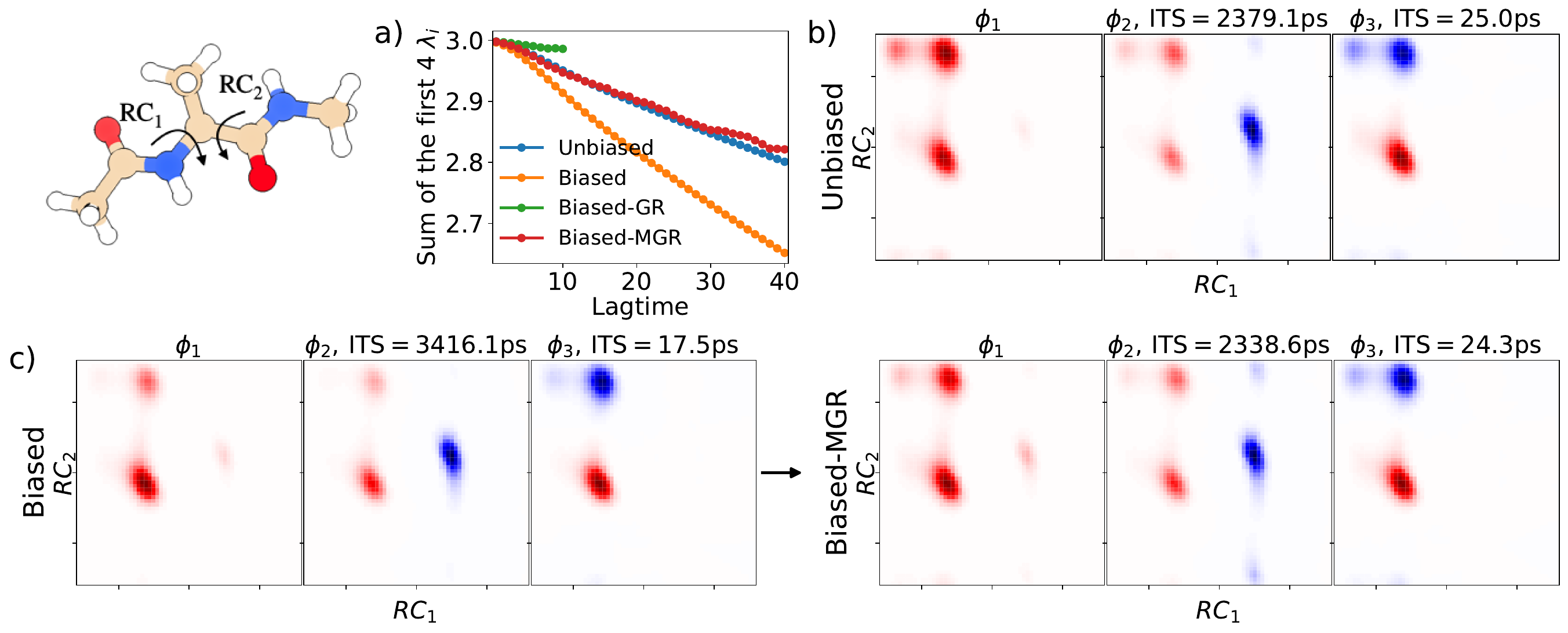}
\end{center}
\vspace{-4mm}
\caption{Results for the 22 atoms Alanine Dipeptide system. \textbf{a)} sum of the first five eigenvalues, reflecting the intrinsic relaxation behavior of the dynamics, as a function of lag time (lagtime unit is $40 \text{fs}$). \textbf{b)} Dominant left eigenfunctions of the transition matrix with corresponding ITS values are shown in the title at lag time $1.6\text{ps}$. Unbiased results serve as reference. \textbf{c)} Dominant left eigenfunctions from the biased trajectory and the MGR-recovered results. GR fails to construct a convergent Markov state model due to excessive variance.}
\label{fig:alanine1}
\end{figure}

We next consider alanine dipeptide \cite{donati2017girsanov}, a widely used benchmark system in molecular dynamics for studying conformational transitions. The backbone torsion angles $\phi$ and $\psi$ are two important reaction coordinates, containing several important metastable basins. Sampling transitions between these basins is challenging due to the presence of high free-energy barriers. To accelerate exploration, we perform biased simulations by introducing umbrella potentials along $\phi$ and $\psi$, which distort the original equilibrium distribution and transition probabilities. Detailed energy function and simulation information can be found in Appendix \ref{appendix:alanine}.

Figure~\ref{fig:alanine1}a) reports the sum of the first five eigenvalues as a function of lag time. MGR consistently follows the unbiased reference, whereas GR shows clear deviations. Due to large weight variance and rapidly collapsing ESS, GR fails to produce a stable transition matrix beyond lag time $10\tau = 400\,\text{fs}$ and thus cannot construct a valid MSM at longer time scales. Figure~\ref{fig:alanine1}b) shows the dominant left eigenfunctions and corresponding ITS at lag time $40\tau = 1.6\,\text{ps}$ under the unbiased trajectory, serving as the evaluation reference. Since GR has already failed at this lag time, only MGR results are reported in Figure~\ref{fig:alanine1}c). Both the eigenfunctions and implied timescales demonstrate that MGR successfully reproduces the unbiased behavior.

\begin{figure}[t!]
\begin{center}
\includegraphics[width=1\textwidth]{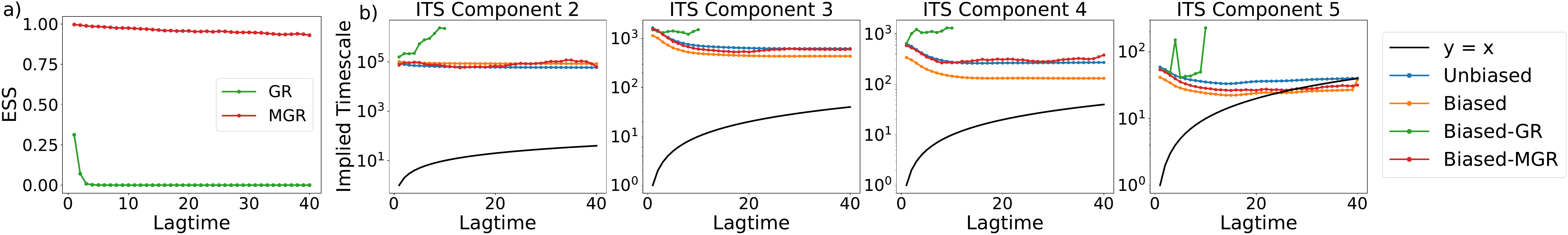}
\end{center}
\vspace{-5mm}
\caption{Supplementary results for alanine dipeptide. \textbf{a)} Relative ESS as a function of 
lag time for GR and MGR. \textbf{b)} First four implied timescales $t_i(\tau)$ for modes 
$i = 2, \ldots, 5$ (log scale). MGR tracks the unbiased ITS closely, while GR exhibits large 
deviations and fails entirely beyond $10\tau = 400\,\text{fs}$.}
\label{fig:alanine-ITS}
\end{figure}

Figure~\ref{fig:alanine-ITS}a) compares the relative ESS as a function of lag time: MGR maintains substantially higher ESS throughout, while GR collapses rapidly. Figure~\ref{fig:alanine-ITS}b) reports the first four individual ITS, where MGR tracks the unbiased reference with markedly less fluctuation. To further illustrate the failure mode of GR, Figure~\ref{fig:alanine-gr} displays the dominant eigenfunctions reweighted by GR at lag times $\tau$ and $10\tau$. At $\tau$, GR yields reasonable corrections, but at $10\tau$ the eigenfunctions become severely distorted and the metastable partitions blur. In contrast, MGR retains stable eigenfunctions across all lag times that closely match the unbiased reference, consistent with its superior ESS and ITS accuracy.

\subsection{Deca-Alanine}

Finally, we validate MGR on deca-alanine (Ala\textsubscript{10}), a 103-atom peptide that presents a substantially higher-dimensional reweighting challenge compared to alanine dipeptide. Simulations are performed using the Amber14 force field \cite{maier2015ff14sb} with the GBn2 implicit solvent model \cite{nguyen2013improved}. In order to accelerated exploration, a harmonic umbrella potential $U(r) = \frac{1}{2}k(r-r_0)^2$ is applied to the end-to-end distance between the two terminal $\text{C}{\alpha}$ atoms, with $k=20 \mathrm{~kJ} \mathrm{~mol}^{-1} \mathrm{~nm}^{-2}$ and $r_0=0.5\mathrm{~nm}$. Both biased and unbiased reference trajectories are propagated for 5 $\mu\mathrm{s}$. Further simulation details are provided in Appendix~\ref{appendix:ala10}.

\begin{figure}[t!]
\begin{center}
\includegraphics[width=1.\textwidth]{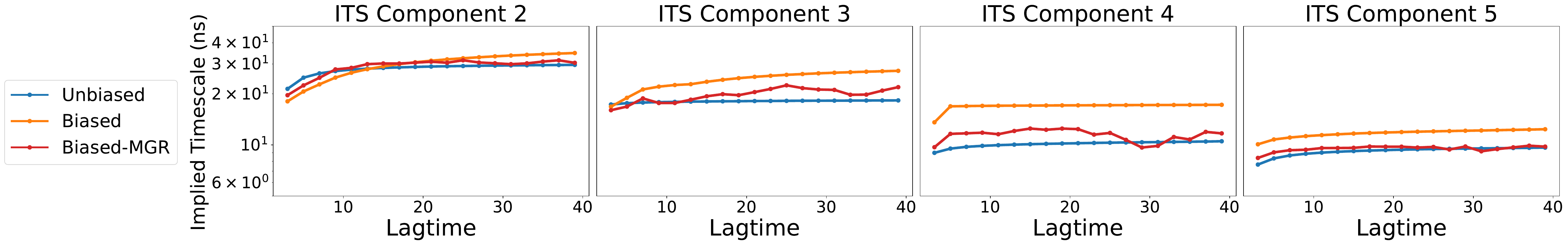}
\end{center}
\vspace{-4mm}
\caption{First four ITS as a function of lag time for the Ala\textsubscript{10} system. Results are shown for the unbiased reference trajectory, the raw biased trajectory, and the biased trajectory reweighted using MGR. MGR-reweighted estimates closely track the unbiased reference across all lag times, whereas the raw biased estimates deviate substantially due to the distortion introduced by the umbrella potential.}
\label{fig:ala10-its}
\end{figure}

\begin{figure}[b!]
\begin{center}
\includegraphics[width=1.\textwidth]{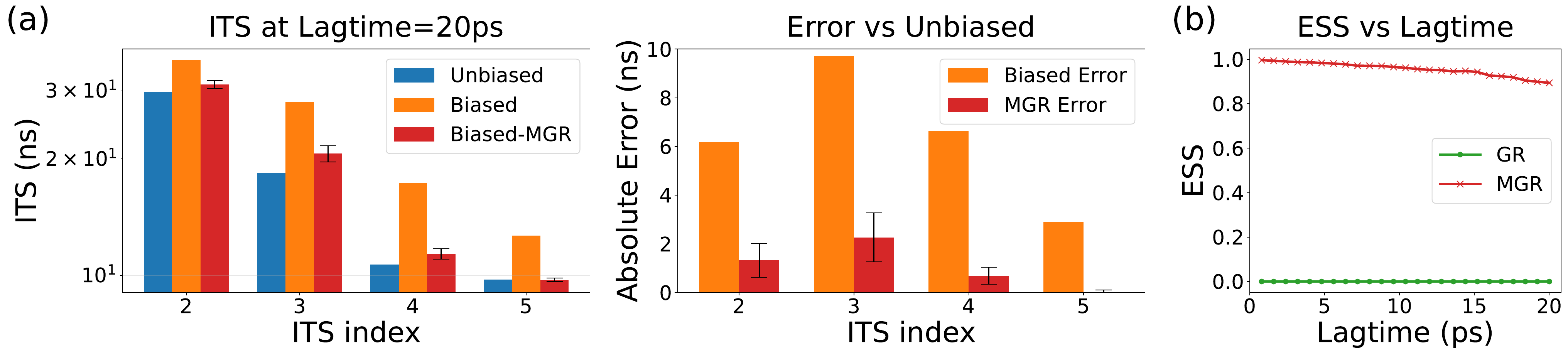}
\end{center}
\vspace{-4mm}
\caption{Results for the Ala\textsubscript{10} system. \textbf{Left:} Mean and standard deviation of individual implied timescales at lag time $\tau = 20\,\mathrm{ps}$ for the unbiased reference, raw biased, and MGR-reweighted trajectories. Error bars represent statistics calculated across 5 independent replicates. MGR recovers quantitative agreement with the unbiased reference. \textbf{Right:} ESS as a function of lag time. MGR maintains substantially higher ESS than GR throughout, demonstrating its superior statistical efficiency in this high-dimensional system.}
\label{fig:ala10-ess}
\end{figure}

Given the high dimensionality of the Ala\textsubscript{10} system, we first perform dimensionality reduction using time-lagged independent component analysis (TICA)~\cite{molgedey1994separation, perez2013identification}, projecting the trajectory onto the leading 20 TICA components. These TICA features are then used both as input for training the MGR model and as the basis for constructing all subsequent MSMs.

Figure~\ref{fig:ala10-its} shows the first four implied timescales as a function of lag time. The ITS obtained from MGR-reweighted biased trajectories closely track the unbiased reference across all lag times, whereas the raw biased estimates deviate markedly, reflecting the distortion introduced by the umbrella potential. Figure~\ref{fig:ala10-ess} provides a direct comparison of the individual ITS values at lag time $\tau=20\mathrm{ps}$, further demonstrating quantitative agreement between MGR and the unbiased reference. It also reports the ESS as a function of lag time. MGR maintains substantially higher ESS than GR throughout, underscoring its practical advantage in high-dimensional systems where the full path-level likelihood ratio incurs prohibitive variance.

These examples demonstrate that MGR achieves consistently strong performance in molecular dynamics. It maintains stable effective sample sizes, produces reliable implied timescales, and accurately recovers equilibrium and kinetic properties, even under large system and at long lag times where traditional GR fails. These results highlight MGR provides a practical framework for reweighting biased simulations in MD.

\section{Conclusion}
Our work introduces Marginal Girsanov Reweighting (MGR), a principled approach for estimating unbiased properties from perturbed paired data. Based on biased pathwise Girsanov reweighting (GR)---which computes the ratio along a specific trajectory---MGR learns a marginal estimator by integrating over intermediate states. This approach mitigates the variance blow-up of GR and can yield a stable estimate of the transition probability under long lags. We show that MGR can be implemented using the adapted classifier-based density ratio estimation, and outperforms the baseline on benchmarks.

Despite MGR’s advantages, challenges remain. When the perturbed drift deviates substantially from the reference dynamics, GR weights become unstable, which in turn hampers MGR training. Careful choice of reaction coordinates is therefore essential in practice. Moreover, while we implemented MGR using standard classifier-based ratio estimation, richer neural estimators and architectures hold promise for further gains. 
In addition, given the rapid development of physical systems and molecular dynamics, an interesting direction is to extend MGR toward simulation-based Bayesian inference and multi-ensemble estimators \cite{zhang2026pi}.
Exploring these directions will broaden the scope of MGR, paving the way for robust, ML-driven reweighting methods applicable across scales, domains, and dynamical systems.

\section*{Acknowledgements}
This work was partially supported by the Wallenberg AI, Autonomous Systems and Software Program (WASP) funded by the Knut and Alice Wallenberg Foundation and by the National Natural Science Foundation of China (NSFC) under grant number 12571463. The research visit of the first author was funded by Tongji University. Preliminary results were enabled by resources provided by the National Academic Infrastructure for Supercomputing in Sweden (NAISS) at Alvis (project: NAISS 2025/22-463), partially funded by the Swedish Research Council through grant agreement no.~2022-06725.

%\section*{Reproducibility Statement}
%All theoretical assumptions are stated explicitly, with complete proofs in the Appendix. Experimental details are described in the main text and Appendix, and the code for data generation and model training is provided in the supplementary materials.

% does not count towards 9 page limit.
\bibliography{iclr2026_conference}
% \printbibliography

\newpage
\appendix
% ——— 附录统一编号：S.1, S.2, ... ———
\makeatletter
\setcounter{figure}{0}
\renewcommand{\thefigure}{S.\arabic{figure}}
\renewcommand{\theHfigure}{S.\arabic{figure}}
\setcounter{table}{0}
\renewcommand{\thetable}{S.\arabic{table}}
\renewcommand{\theHtable}{S.\arabic{table}}
\setcounter{equation}{0}
\renewcommand{\theequation}{S.\arabic{equation}}
\renewcommand{\theHequation}{S.\arabic{equation}}
\makeatother

\section{Girsanov Reweighting for Underdamped Langevin Dynamics}\label{appendix:underdamped}
In molecular dynamics simulations, the system is typically propagated by underdamped Langevin dynamics rather than the overdamped discussed in Section~\ref{sec:gr}. The equations of motion for position $x_t \in \mathbb{R}^d$ and momentum $v_t \in \mathbb{R}^d$ are
\begin{align}
    \mathrm{d}x_t &= v_t\,\mathrm{d}t, \nonumber \\
    \mathrm{d}v_t &= -\nabla V(x_t)\,\mathrm{d}t - \gamma\, v_t\,\mathrm{d}t + \sigma\,\mathrm{d}W_t, \label{eq:underdamped-v}
\end{align}
where $V(x)$ is the potential energy, $\gamma > 0$ is the friction coefficient, $\sigma = \sqrt{2\gamma k_B T}$ satisfies the fluctuation-dissipation relation, and $W_t$ is a standard $d$-dimensional Wiener process. The corresponding biased dynamics replaces $V(x)$ with a modified potential $\tilde{V}(x, t) = V(x) + U(x, t)$, where $U$ is the bias potential introduced by the enhanced sampling method.

Since the stochastic noise enters only through the momentum equation~\ref{eq:underdamped-v}, the Girsanov change of measure applies to the momentum path. For a trajectory segment $\{(x^k, v^k)\}_{k=0}^{N}$ with time step $\Delta t$, the log-weight takes the form~\cite{kieninger2021path,kieninger2023girsanov}
\begin{align}
    \log w_{\tau}^{\mathrm{GR}} = \sum_{k=0}^{N-1} \left( \frac{\nabla U(x^k, t)^{\top}}{\sigma} \sqrt{\Delta t}\,\eta^k - \frac{\Delta t}{2} \left\| \frac{\nabla U(x^k, t)}{\sigma} \right\|^2 \right), \label{eq:gr-underdamped}
\end{align}
where $\sqrt{\Delta t}\,\eta^k$ is the Wiener increment in the momentum update step.

The precise form of the Wiener increment $\eta^k$ depends on the Langevin integrator. For the ABOBA splitting scheme~\cite{leimkuhler2013rational} used in this paper, which is widely adopted in modern MD packages, a single integration step consists of the following sub-steps:
\begin{align*}
    &\text{(A)}\quad & v^{k+\frac{1}{4}} &= v^k + \tfrac{\Delta t}{2}\,a^k, \\
    &\text{(B)}\quad & x^{k+\frac{1}{2}} &= x^k + \tfrac{\Delta t}{2}\,v^{k+\frac{1}{4}}, \\
    &\text{(O)}\quad & v^{k+\frac{3}{4}} &= e^{-\gamma \Delta t}\,v^{k+\frac{1}{4}} + \sqrt{\frac{k_B T}{m}\left(1-e^{-2\gamma \Delta t}\right)}\;\eta^k, \\
    &\text{(B)}\quad & x^{k+1} &= x^{k+\frac{1}{2}} + \tfrac{\Delta t}{2}\,v^{k+\frac{3}{4}}, \\
    &\text{(A)}\quad & v^{k+1} &= v^{k+\frac{3}{4}} + \tfrac{\Delta t}{2}\,a^{k+1}, 
\end{align*}
where $a^k = -\nabla \tilde{V}(x^k, t)$ is the acceleration under the biased potential, and $\eta^k \sim \mathcal{N}(0, I_d)$ is the random vector drawn in the O~step. The random numbers $\eta^k$ can be recorded during the simulation, and the reweighting factor in Eq.~\ref{eq:gr-underdamped} can be accumulated on-the-fly at acceptable computational overhead.
 
In this work, we use the OpenMM implementation of GR~\cite{schafer2024implementation}, which records the necessary random numbers and forces at each integration step and computes the Girsanov weights as part of the simulation workflow.

\section{Variance of Girsanov Reweighting} \label{appendix:gr-variance}
We analyze the variance behavior of Girsanov reweighting over a continuous time interval $[t, t+\tau]$. According to Girsanov theory \cite{girsanov1960transforming}, the log-weight under the Girsanov transformation can be expressed as
\[
\log w_\tau^{\mathrm{GR}}\left(\mathbf{x}_{t, \tau}\right)=\int_t^{t+\tau} u\left(x_s, s\right)^{\top} \mathrm{d} W_s-\frac{1}{2} \int_t^{t+\tau}\left\|u\left(x_s, s\right)\right\|^2 \mathrm{~d} s,
\]
where $u(x_t, t) := \frac{\nabla U(x_t, t)}{\sigma}$ denotes the rescaled drift difference.

Given a fixed control path $\mathbf u_{t, \tau}:=\{ u(x_s, s)\}_{s=t}^{t+\tau}$, the expectation and variance of the log-weight are
\begin{align*}
& \mathbb{E}\left[\log  w_\tau^{\mathrm{GR}}\left(\mathbf{x}_{t, \tau}\right) \mid \mathbf u_{t, \tau}\right]=-\frac{1}{2} \int_t^{t+\tau}\left\|u(x_s, s)\right\|^2 \mathrm d s, \\
& \operatorname{Var}\left( \log w_\tau^{\mathrm{GR}}\left(\mathbf{x}_{t, \tau}\right)\mid \mathbf u_{t, \tau}\right)=\int_t^{t+\tau}\left\|u(x_s, s)\right\|^2 \mathrm d s .
\end{align*}
We now consider a time-discretized version of the variance over an interval $[t, t+\tau]$ with $N$ steps of size $\Delta t = \tau/N$. Denote the discretized control as $\mathbf u^{0:N-1}_{t,\tau} := \{u(x^{k}, t^k)\}_{k=0}^{N-1}$. Exponentiating this log-weight, the conditional variance of the log-weight becomes
\[
\operatorname{Var}\left( w_\tau^{\mathrm{GR}}\left(\mathbf{x}_{t, \tau}^{0:N}\right) \mid \mathbf u^{0:N-1}_{t,\tau}\right)=\exp \left(\sum_{k=0}^{N-1}\left\|u(x^{k}, t^k)\right\|^2 \Delta t \right)-1.
\]
This reveals that the variance of the Girsanov weight grows exponentially with both trajectory duration $\tau$ and the dimension of control magnitude $u$, where $\|u(\cdot, t) \|^2$ scales with the dimension $d$.

\section{Ratio of transition density} 
\label{appendix:transition}
Let $\mathbf x_{t,\tau} = \{x_s \}_{s=t}^{t+\tau}$ denote the path space of continuous trajectories. We consider two probability measures, which are $\mu (\mathbf x_{t,\tau})$ (the original process defined in Eq.~\ref{eq:unbiasedsde}) and $\tilde \mu (\mathbf x_{t,\tau})$ (the perturbed process with drift term $\tilde f(\cdot, t)$). For any bounded measurable test function $O:\mathbb R^d\times \mathbb R^d \to \mathbb R$,
\begin{eqnarray*}
    \mathbb E_{\rho_{\tau}(x_t, x_{t+\tau})}\left[O(x_t, x_{t+\tau}) \right]& = &\int O(x_t, x_{t+\tau})\rho_{\tau}(x_t, x_{t+\tau})\mathrm d x_t \mathrm d x_{t+\tau} \\
    &=&\int O(x_t, x_{t+\tau})\delta(X_t-x_t)\delta(X_{t+\tau}-x_{t+\tau})\mathrm d \mu (\mathbf x_{t,\tau}) \\
    &=&\int O(x_t, x_{t+\tau})\delta(X_t-x_t)\delta(X_{t+\tau}-x_{t+\tau})\frac{\mathrm d \mu}{\mathrm d \tilde \mu}( \mathbf x_{t,\tau}) \mathrm d \tilde \mu (\mathbf x_{t,\tau}) \\
    &=&\int O(x_t, x_{t+\tau})\tilde \rho_{\tau}(x_t, x_{t+\tau}) \frac{\mathrm d \mu}{\mathrm d \tilde \mu}( \mathbf x_{t,\tau}) \mathrm d \tilde \mu (\mathbf x_{t,\tau}) \\
    &:=& \mathbb E_{\tilde \rho_{\tau}(x_t, x_{t+\tau})}\left[w_{\tau}(x_t, x_{t+\tau}) O(x_t, x_{t+\tau}) \right],
\end{eqnarray*}
where $w_{\tau}(x_t, x_{t+\tau}) = \mathbb E_{\tilde \mu} \left[\frac{\mathrm d \mu}{\mathrm d \tilde \mu}( \mathbf x_{t,\tau}) \mid X_t = x_t, X_{t+\tau} = x_{t+\tau} \right]$. Here, $\rho_{\tau}(x_t, x_{t+\tau})$, $\tilde \rho_{\tau}(x_t, x_{t+\tau})$ denote the joint marginal distributions under the original and perturbed processes respectively.

This formulation suggests that the ratio of transition densities $w_{\tau}(x_t, x_{t+\tau})$ can be estimated by Girsanov reweighting. However, beyond the well-known issue of rapidly growing variance in Appendix~\ref{appendix:gr-variance}, Girsanov reweighting computes weights tied to specific trajectories $\mathbf x_{t, \tau}$, whereas the desired transition ratio $w_{\tau}(x_t, x_{t+\tau})$ corresponds to an expectation over paths connecting the given endpoints.

\section{Consistency of the approximate $\rho_{k\tau}(x, y)$} \label{appendix:iterativetraining}
For the given iteration $k>1$, let $Z:=X_{t+(k-1)\tau}$ and denote the short lag-$\tau$ path $\mathbf x_{t+(k-1)\tau,\tau} =\{x_s \}_{s=t+(k-1)\tau}^{t+k\tau}$. For any bounded measurable $O: \mathbb{R}^d \times \mathbb{R}^d \rightarrow \mathbb{R}$, an unbiased estimation of transition properties can be obtained by 
\begin{eqnarray*}
\mathbb E_{\rho_{k\tau}(x_t, x_{t+k\tau})} \left[O(x_t, x_{t+k\tau})) \right]
&=&\mathbb E_{\tilde\rho_{(k-1)\tau}(x_t, z),\ \tilde \mu(\mathbf x_{t+(k-1)\tau,\tau}\mid z)}\left[c_tO(x_t, x_{t+k\tau}) \right],
\end{eqnarray*}
where $c_t=w_{(k-1)\tau}(x_t, z) w_{\tau}^{\text{GR}}(\mathbf x_{t+(k-1)\tau,\tau})$. Here, $w_{(k-1)\tau}(x_t, z)$ is the marginal weights inherited from the previous iteration, and $w_{\tau}^{\text{GR}}(\mathbf x_{t+(k-1)\tau,\tau})$ is the pathwise Girsanov reweighting introduced in Section~\ref{sec:gr}. In practical biomolecular applications, choosing $O(x_t, x_{t+k\tau}) = \mathbf{1}_{B_i}(x_t)\,\mathbf{1}_{B_j}(x_{t+k\tau})$ yields the unbiased lag-$k\tau$ transition counts between metastable states $B_i$ and $B_j$ from which Markov state models at longer lag times can be constructed according to Eq.~\ref{eq:msm_count}.

In the algorithm, we approximate this expectation by Monte Carlo. Extract all paired data $\{(x^i, \mathbf x_{i+(k-1)\tau,\tau}^{0:N})\}_{i=1}^M$ under $\tilde\mu$ from either a single long trajectory or multiple trajectories, where each short segment $\mathbf x_{i+(k-1)\tau,\tau}^{0:N}$ connects the intermediate endpoints $(z^i, y^i)$ with lag $\tau$, (thus the total lag between $x^i$ and $y^i$ is $k\tau$). With an assigned pathwise weight $c^i = w_{(k-1)\tau}(x^i, z^i) w_{\tau}^{\text{GR}}(\mathbf x_{i+(k-1)\tau,\tau}^{0:N}))$, the expectation can be estimated by
\[
\mathbb E_{\rho_{k\tau}(x_t, x_{t+k\tau})}\left[O(x_t, x_{t+k\tau}) \right] \approx \sum_{i=1}^M \frac{c^i O(x^i, y^i)}{\sum_{i=1}^{M} c^i}.
\]
The estimation error approaches zero when $w_{(k-1)\tau}$ is accurate and $M\rightarrow \infty$.

\textit{Proof:}

By Chapman–Kolmogorov \cite{hachigian1963collapsed}, we have
\begin{align}
\mathbb E_{\rho_{k\tau}(x_t, x_{t+k\tau})}\left[O(x_t, x_{t+k\tau}) \right] =& \int \rho_{k\tau}(x_t, x_{t+k\tau})O(x_t, x_{t+k\tau}) \mathrm d x_t \mathrm d x_{t+k\tau} \nonumber \\
=& \int \rho_{(k-1)\tau}(x_t, z) p(x_{t+k\tau}\mid z)O(x_t, x_{t+k\tau}) \mathrm d z \mathrm d x_t \mathrm d x_{t+k\tau} \nonumber \\
=& \int \frac{\rho_{(k-1)\tau}(x_t, z)p( x_{t+k\tau}\mid z)}{\tilde\rho_{(k-1)\tau}(x_t, z)\tilde p( x_{t+k\tau}\mid z)} \cdot \nonumber \\
&\quad \tilde\rho_{(k-1)\tau}(x_t, z)\tilde p( x_{t+k\tau}\mid z) O(x_t, x_{t+k\tau}) \mathrm d z \mathrm d x_t \mathrm d x_{t+k\tau}. \nonumber
\end{align}

According to Section~\ref{sec:setting}, $\frac{p( x_{t+k\tau}\mid z)}{\tilde p( x_{t+k\tau}\mid z)} = \mathbb E_{\tilde\mu(\mathbf x_{t+(k-1)\tau,\tau})} \left[ w_{\tau}^{\text{GR}}(\mathbf x_{t+(k-1)\tau,\tau})\mid Z=z, X_{t+k\tau}=x_{t+k\tau} \right]$ and $w_{(k-1)\tau}(x_t, z)=\frac{\rho_{(k-1)\tau}(x_t, z)}{\tilde \rho_{(k-1)\tau}(x_t, z)}$. It yields
\begin{align*}
    \mathbb E_{\rho_{k\tau}(x_t, x_{t+k\tau})}\left[O(x_t, x_{t+k\tau}) \right] =& \int \tilde\rho_{(k-1)\tau}(x_t, z)\tilde p( x_{t+k\tau}\mid z) w_{(k-1)\tau}(x_t, z) w_{\tau}^{\text{GR}}(\mathbf x_{t+(k-1)\tau,\tau})  \nonumber \\
    &\quad O(x_t, x_{t+k\tau})  \tilde \mu(\mathrm d \mathbf x_{t+(k-1)\tau,\tau}\mid z, x_{t+k\tau})\mathrm d z \mathrm d x_t \mathrm d x_{t+k\tau} \\
    =& \int \tilde\rho_{(k-1)\tau}(x_t, z) \tilde \mu(\mathrm d \mathbf x_{t+(k-1)\tau,\tau}\mid z) \\
    &\quad w_{(k-1)\tau}(x_t, z) w_{\tau}^{\text{GR}}(\mathbf x_{t+(k-1)\tau,\tau})O(x_t, x_{t+k\tau}) \mathrm d z \mathrm d x_t \nonumber \\
    :=& \mathbb E_{\tilde\rho_{(k-1)\tau}(x_t, z),\ \tilde \mu(\mathbf x_{t+(k-1)\tau,\tau}\mid z)}\left[c_tO(x_t, x_{t+k\tau}) \right].
\end{align*}

Due to the fact $\mathbb E_{\rho_{k\tau}(x_t, x_{t+k\tau})}\left[O(x_t, x_{t+k\tau}) \right]= \mathbb E_{\tilde\rho_{k\tau}(x_t, x_{t+k\tau})}\left[ w_{k\tau}(x_t, x_{t+k\tau}) O(x_t, x_{t+k\tau}) \right]$, we can also prove $w_{k\tau}(x_t, x_{t+k\tau}) = \mathbb E_{\tilde\rho_{(k-1)\tau}(x_t, z),\ \tilde \mu(\mathbf x_{t+(k-1)\tau,\tau}\mid z)}\left[c_t\right]$, which provides the theoretical consistency across all lag scales.

Let $\{(x^i, y^i)\}_{i=1}^M$ be endpoint pairs sampled under $\tilde \mu$ and their pathwise weights drawn from $c^i = w_{(k-1)\tau}(x^i, z^i) w_{\tau}^{\text{GR}}(\mathbf x_{i+(k-1)\tau,\tau}^{0:N})$. 
By the strong law of large numbers for Markov chains with ergodic assumption \cite{breiman1960strong},
\begin{align*}
\sum_{i=1}^M \frac{c^i O(x^i, y^i)}{\sum_{i=1}^{M} c^i}\xrightarrow{\text { a.s. }}&\mathbb E_{\tilde\rho_{(k-1)\tau}(x_t, z),\ \tilde \mu(\mathbf x_{t+(k-1)\tau,\tau}\mid z)}\left[c_tO(x_t, x_{t+k\tau}) \right] = \mathbb E_{\rho_{k\tau}(x_t, x_{t+k\tau})} \left[O(x_t, x_{t+k\tau})) \right].
\end{align*}

\section{Training Algorithm}

\begin{algorithm}[b!]
\caption{Marginal Girsanov Reweighting (MGR)}
\label{alg:MGR}
\begin{algorithmic}[1]
\Require Simulation trajectory $\{x_t\}_{t=0}^{T}$ from perturbed dynamics, lagtime $\tau$, the maximum training iteration $K$, learning rate $\eta$.
\For{$k=1$ to $K$}
\Statex
\textit{\textbf{Step 1:}} \textbf{Collect paired data}
\State Collect training pairs $\{(x_t, x_{t+k\tau})\}_{t=0}^{T-k\tau}$;

\Statex
\textit{\textbf{Step 2:}} \textbf{Compute pathwise weights}

\State Compute Girsanov weight $w_{\tau}^{\text{GR}}(\mathbf x_{t+(k-1)\tau,\tau}^{0:N})$ using Eq.~\ref{eq:gr-discrete};
\State Compute pathwise weight:
\If{$k=1$}
    \State $c_t = w_{\tau}^{\text{GR}}(\mathbf x_{t+(k-1)\tau,\tau}^{0:N})$;
\Else
    \State $c_t = w_{(k-1)\tau}(x_t, x_{t+(k-1)\tau})w_{\tau}^{\text{GR}}(\mathbf x_{t+(k-1)\tau,\tau}^{0:N})$;
\EndIf

\Statex
\textit{\textbf{Step 3:}} \textbf{Train density ratio estimator}
\For{each training epoch}
    \For{each minibatch $\{(x^{(i)}, y^{(i)}, c^{(i)})\}_{i=1}^{B}$ drawn from $\{(x_t, x_{t+k\tau}, c_t)\}_{t=0}^{T-k\tau}$}
        \State Compute weighted binary classification loss with normalized weight $c^{(i)}$:
        \[
        \mathcal{L}(\theta) = -\frac{1}{B} \sum_{i=1}^{B} \left[ c^{(i)} \log h_{\theta}(x^{(i)}, y^{(i)}) + \log (1 - h_{\theta}(x^{(i)}, y^{(i)})) \right];
        \]
        \State Update parameters: $\theta \gets \theta - \eta \cdot \nabla_\theta \mathcal{L}(\theta)$;
    \EndFor
\EndFor
\State Update the marginal ratio $w_{k\tau}(x_t, x_{t+k\tau}) = \frac{h_{\theta}^*(x_t,x_{t+k\tau})}{1-h_{\theta}^*(x_t,x_{t+k\tau})}$;
\EndFor
\State \Return Marginal ratio $w_{k\tau}(x_t, x_{t+k\tau})$ by model
\end{algorithmic}
\end{algorithm}

To summarize the MGR procedure, we present the full training and evaluation process in Algorithm~\ref{alg:MGR}. At each iteration indexed by lag time $k\tau$, the goal is to train a classifier to estimate the marginal density ratio $w_{k\tau}(x_t, x_{t+k\tau})$ between the original and perturbed transition densities. This is achieved by constructing pathwise weights $c_t$ that combine the short-time Girsanov weight $w_\tau^{\mathrm{GR}}$ with the model prediction from the previous iteration $w_{(k-1)\tau}$. The resulting $w_{k\tau}$ is then used in the next iteration, allowing the model to progressively extend from short to long lag times.

\section{Experimental Details} \label{appendix:experiments}

% \subsection{MSMs construction}\label{appendix:rc-input}

% In the experiments on alanine dipeptide and deca-alanine, we train the MGR model and construct MSMs on a set of pre-selected reaction coordinates $z_t = \xi(x_t)$ rather than the full atomic configuration $x_t$. For alanine dipeptide, $z_t$ consists of the two backbone dihedral angles $(\phi, \psi)$. For deca-alanine, $z_t$ is a $20$-dimensional representation obtained by TICA~\cite{perez2013identification} on the backbone dihedral features and pairwise $C_{\alpha}$ distances between residues separated by more than five residues along the chain. Under the framework of optimal reaction coordinates~\cite{bittracher2023optimal}, this choice is reasonable and yields good performance in our experiments. A systematic study of how the quality of the reaction coordinate affects the accuracy of the MGR reweighting factor is an interesting direction for future work.

In the experiments on Four Well and Muller Brown systems, we train the MGR model and construct MSMs in the coordinates space. For alanine dipeptide and deca-alanine systems, a set of pre-selected reaction coordinates $z_t = \xi(x_t)$ are used to train the MGR model and construct MSMs, as we found this to yield better performance in practice. For alanine dipeptide, $z_t$ consists of the two backbone dihedral angles $(\phi, \psi)$. For deca-alanine, $z_t$ is a $20$-dimensional representation obtained by TICA~\cite{perez2013identification} on the backbone dihedral features and pairwise $C_{\alpha}$ distances between residues separated by more than five residues along the chain.

\subsection{1 Dimensional Four Well} \label{appendix:fourwell}

We consider an overdamped Langevin dynamics on a one–dimensional four–well landscape,
\[
    \mathrm{d} X_t=-\nabla V\left(X_t\right) \mathrm{d} t+ \sigma \mathrm{~d} W_t,
\]
where $V(x)=4\left(x^8+0.8 e^{-80 x^2}+0.2 e^{-80(x-0.5)^2}+0.5 e^{-40(x+0.5)^2}\right)$, $\sigma=1$.

\begin{figure}[h!]
\vspace{-3mm}
\begin{center}
\includegraphics[width=1.0\textwidth]{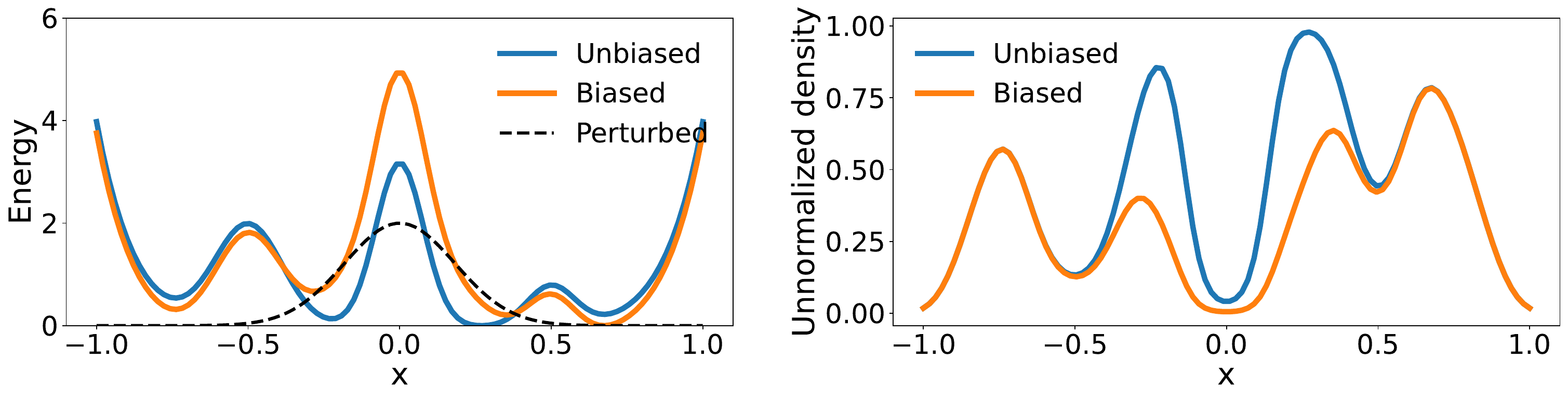}
\end{center}
\vspace{-6mm}
\caption{Energy and density of 1D four-well system. \textbf{Left:} Energy profiles of unbiased potential $V(x)$, biased potential $\widetilde V(x)=V(x)+U(x)$, and the bias term $U(x)$. \textbf{Right:} Theoretical unnormalized probability of the system.}
\label{fig:four-well-energy}
\end{figure}

To accelerate barrier crossings between the two intermediate wells, we introduce a perturbed potential
\[
    \mathrm{d} X_t=-\nabla \widetilde V \left(X_t\right) \mathrm{d} t+ \sigma \mathrm{~d} W_t,
\]
where $\widetilde V (x) = V(x)+U(x),\ U(x)=2e^{-kx^2},\ k=15$. Illustrations of the potentials and stationary densities are provided in Figure~\ref{fig:four-well-energy}.

We use the Euler–Maruyama scheme with time step $\Delta t=10^{-3}$. Each run is integrated up to $T=10,000$, yielding $T/\Delta t = 10^7$ samples per trajectory. We initialize at $X_0 = 0$ and simulate a single trajectory. During the perturbed run, we record the discrete time Girsanov log-weights according to Eq.~\ref{eq:gr-discrete}, i.e.,
\[
\log w_{\Delta t}^{\text{GR}}(x) = -\nabla U(x)\sqrt{\Delta t}\ \xi - \frac{\Delta t}{2}\left(\nabla U(x)\right)^2,
\]
where $\xi$ is the corresponding noise in the simulation. 

We set $\tau=50\Delta t$ as the reference short GR lagtime during training. At this short lagtime, the Girsanov weights are numerically stable, with $\text{ESS}\approx 0.43$. Guidelines for selecting a suitable short lag $\tau$ are provided in Appendix~\ref{appendix:lag-ablation}. In each training iteration $k$, we normalize the pathwise weights
\begin{eqnarray}
c_t = \frac{w_{(k-1)\tau}(x_t,x_{t+(k-1)\tau})\,w_{\tau}^{\mathrm{GR}}(\mathbf x_{t+(k-1)\tau,\tau}^{0:N})}{\frac{1}{M}\sum_t \left( w_{(k-1)\tau}(x_t,x_{t+(k-1)\tau})\,w_{\tau}^{\mathrm{GR}}(\mathbf x_{t+(k-1)\tau,\tau}^{0:N}) \right)}, \label{eq:ct-normalized}
\end{eqnarray}
where $M$ is the number of total paired data in our training dataset. After training, we obtain the estimated marginal ratio \(w_{k\tau}(x_t,x_{t+k\tau})=\frac{h_{\theta}^*(x_t,x_{t+k\tau})}{1-h_{\theta}^*(x_t,x_{t+k\tau}) }\) for reweighting.

\begin{figure}[t!]
\begin{center}
\includegraphics[width=1\textwidth]{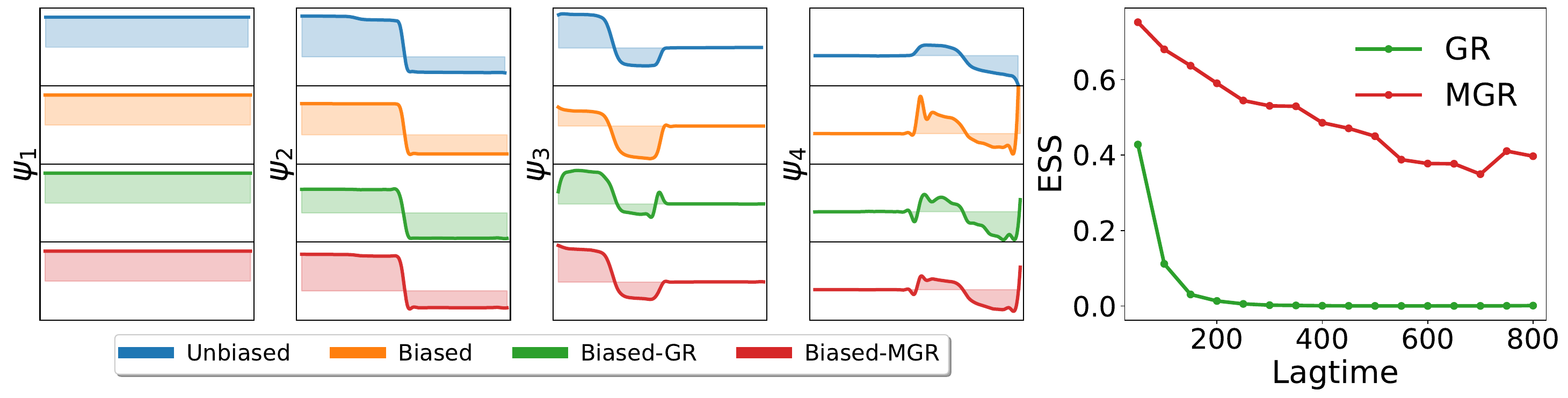}
\end{center}
\vspace{-6mm}
\caption{\textbf{Left:} Dominant right eigenfunctions ($\{\boldsymbol{\psi}_i\}$) of the 
transition matrix at lag time $300\Delta t$. \textbf{Right:} Relative ESS as a function of lag 
time for GR and MGR.}
\label{fig:four-well1}
\end{figure}

\subsection{Müller-Brown Potential}\label{appendix:muller-brown}

\begin{figure}[b!]
\begin{center}
\includegraphics[width=1\textwidth]{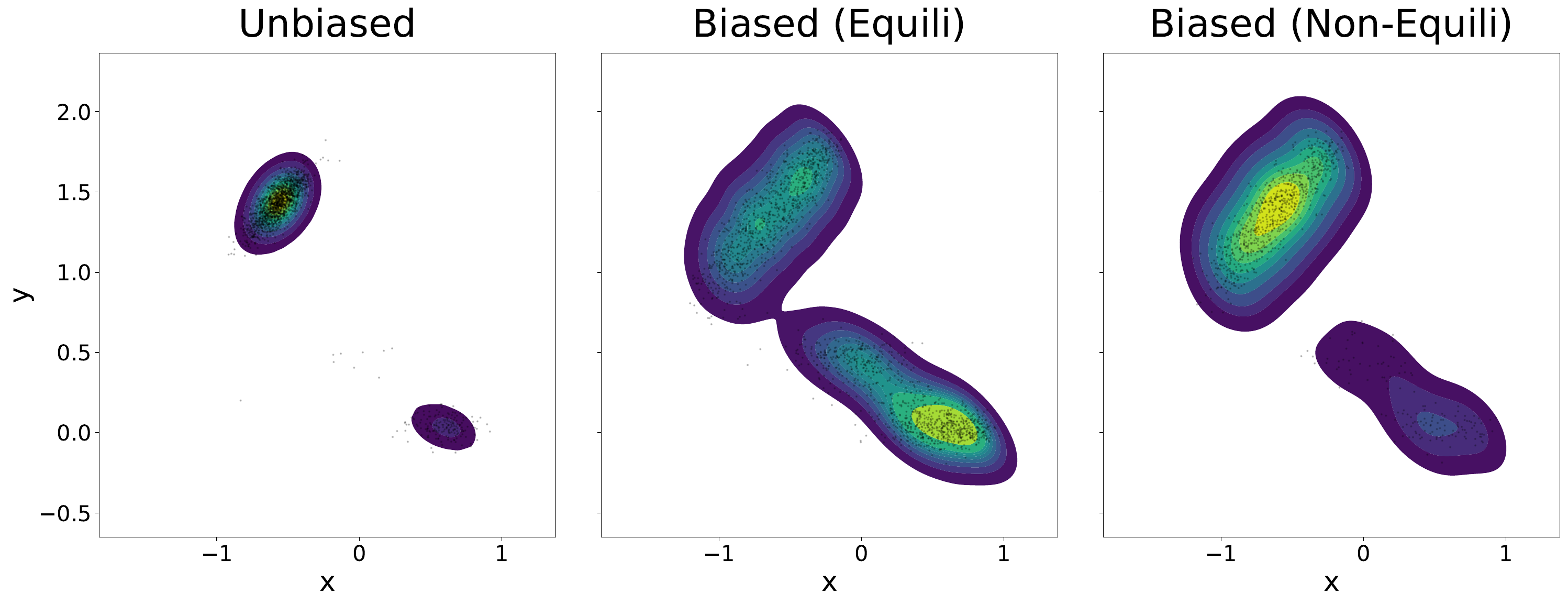}
\end{center}
\vspace{-4mm}
\caption{Müller--Brown potential energy surface and sample distributions. \textbf{Left:} Unbiased simulation. \textbf{Center:} Biased simulation in the equilibrium regime ($10^7$ steps). \textbf{Right:} Biased simulation in the non-equilibrium regime ($10^6$ steps). Samples are overlaid on the potential energy contours.}
\label{fig:mb-potential}
\end{figure}

The Müller--Brown potential is defined as
\begin{equation}
    V(x, y) = \sum_{k=1}^{4} A_k \exp\!\left[a_k(x - x_k^0)^2 + b_k(x - x_k^0)(y - y_k^0) 
    + c_k(y - y_k^0)^2\right],
\end{equation}
with standard parameters as given in~\cite{muller1979location}. All trajectories are initialized from the local minimum at $(0.5,\, 0)$. The overdamped Langevin dynamics are discretized via the Euler-Maruyama scheme with time step $\Delta t = 10^{-3}$ and temperature $T = 1$.

\begin{figure}[t!]
\begin{center}
\includegraphics[width=1\textwidth]{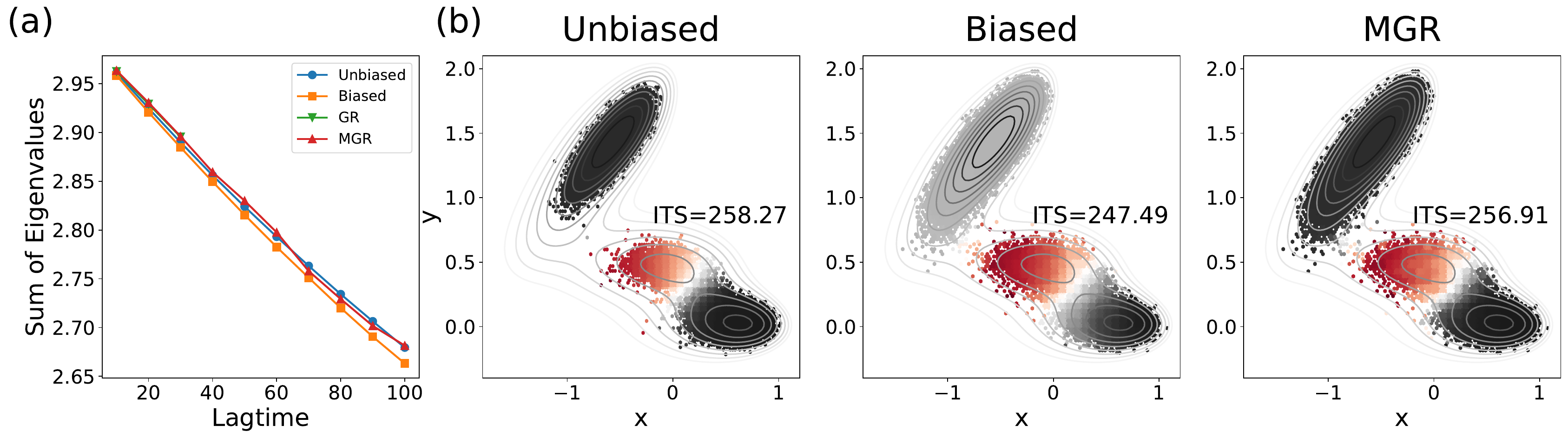}
\end{center}
\vspace{-4mm}
\caption{Müller-Brown potential: non-equilibrium regime ($10^6$ steps). Layout identical to Figure~\ref{fig:mb-eq}. Despite incomplete equilibration, MGR recovers the unbiased eigenvalue 
spectrum and eigenfunctions.}
\label{fig:mb-neq}
\end{figure}

Figure~\ref{fig:mb-potential} shows the potential energy surface together with the sample distributions for the three settings. The unbiased trajectory concentrates in the two principal minima with rare barrier crossings. The equilibrium biased trajectory achieves broad, uniform coverage of the state space, while the non-equilibrium biased trajectory exhibits visibly non-uniform sampling due to incomplete relaxation. Figure~\ref{fig:mb-neq} shows the results obtained from non-equilibrium biased simulations.

\subsection{Alanine Dipeptide}\label{appendix:alanine}
We performed all-atom MD simulations of acetyl-alanine-methylamide (Ac-A-NHMe, alanine dipeptide) in implicit water. The simulation was carried out with the \textsc{OpenMM}~8.2 simulation package \cite{eastman2023openmm} at $300 \text K$. The system employed the Amber14 force field with OBC2 implicit water (``\texttt{amber14-all.xml}'', ``\texttt{implicit/obc2.xml}''). Dynamics were propagated with an Underdamped Langevin integrator with time step $2\text{fs}$. The aggregated simulation time was $1 \mu \text{s}$. Coordinates and Girsanov reweighting (GR) factors were saved every 20 steps ($40 \text{fs}$) using the Girsanov-enabled \textsc{OpenMM} implementation \cite{schafer2024implementation}.

\begin{figure}[h!]
\vspace{-1mm}
\begin{center}
\includegraphics[width=1.0\textwidth]{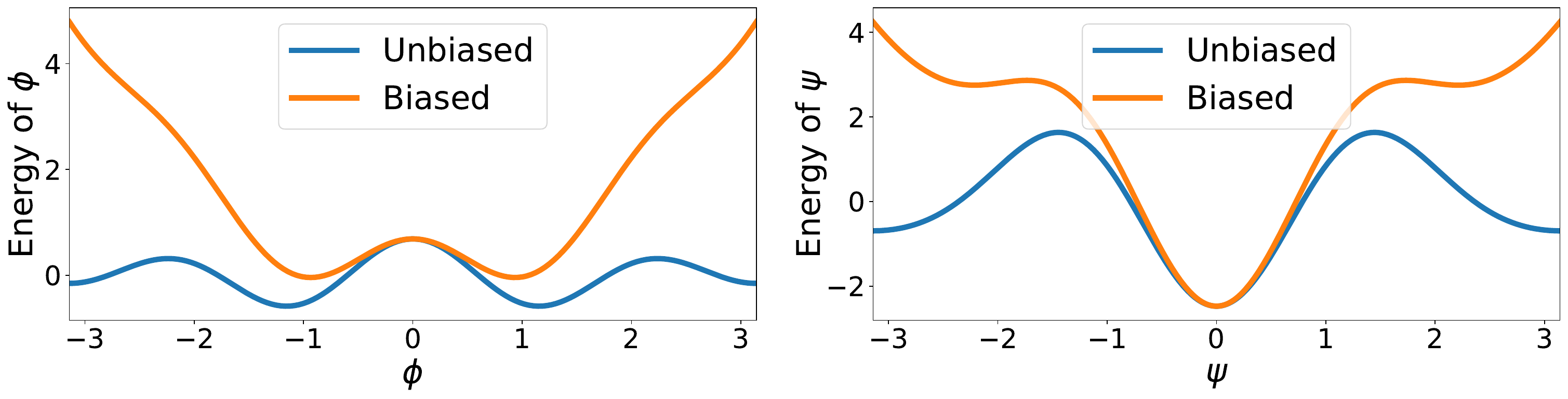}
\end{center}
\vspace{-6mm}
\caption{Alanine dipeptide dihedral energy function. \textbf{Left:} $\phi$ torsion: unbiased energy $V(\phi)$ and biased function $V(\phi)+\frac{1}{2} \kappa_\phi \phi^2$. \textbf{Right:} $\psi$ torsion: unbiased energy $V(\psi)$ and biased function $V(\psi)+\frac{1}{2} \kappa_\psi \psi^2$.}
\label{fig:alanine-energy}
\vspace{-2mm}
\end{figure}

\begin{figure}[t!]
\begin{center}
\includegraphics[width=1\textwidth]{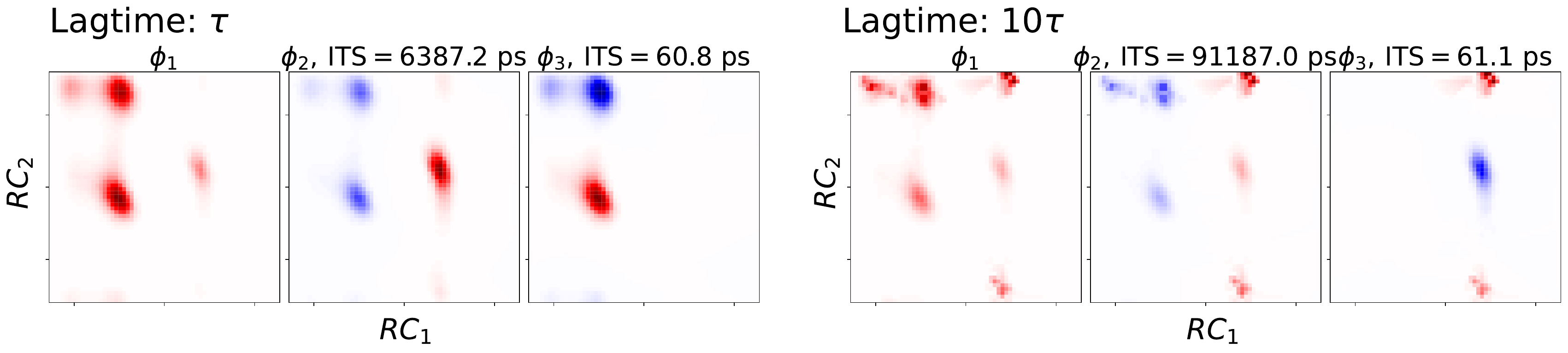}
\end{center}
\vspace{-6mm}
\caption{Dominant eigenfunctions of the biased trajectory reweighted by GR at lag times $\tau$ 
and $10\tau$. At $\tau$, GR produces reasonable corrections; at $10\tau$, the eigenfunctions 
become distorted and metastable partitions are no longer resolved.}
\label{fig:alanine-gr}
\end{figure}

For enhanced sampling, we applied a dihedral bias to the backbone torsions $\phi$ and $\psi$. The unbiased dihedral potentials were
\begin{eqnarray*}
V(\phi)&=&0.27 \cos (2 \phi)+0.42 \cos (3 \phi), \\
V(\psi)&=&0.45 \cos (\psi-\pi)+1.58 \cos (2 \psi-\pi)+0.44 \cos (3 \psi-\pi),
\end{eqnarray*}
and the perturbation was a quadratic restraint
\[
U(\phi, \psi)=\frac{1}{2} \kappa_\phi \phi^2 + \frac{1}{2} \kappa_\psi \psi^2, \quad \kappa_\phi=\kappa_\psi=1,
\]
so that the biased potential is $\widetilde V(\cdot) = V(\cdot)+U(\phi, \psi)$. The perturbation lowers the energy barrier in targeted regions of the $(\phi, \psi)$ free-energy surface, thereby facilitating transitions among metastable basins (Figure~\ref{fig:alanine-energy}), but it also distorts equilibrium and kinetics.

We set $\tau=40\text{fs}$ as the reference short GR lagtime during training. At this short lagtime, the Girsanov weights are numerically stable, with $\text{ESS}\approx 0.31$. In each training iteration $k$, we normalize the pathwise weights according to Eq.~\ref{eq:ct-normalized}.

The reweighted transition counts yield an approximation of the unbiased transition matrix, enabling standard MSM analysis. For alanine dipeptide, we construct the MSM in the key reaction coordinates $(\phi, \psi)$ space. In our model, we likewise use only two angles as inputs to the ratio estimator, which yields satisfactory performance.

Figure~\ref{fig:alanine-gr} displays the dominant eigenfunctions reweighted by GR at lag times $\tau$ and $10\tau$. At $\tau$, GR yields reasonable corrections, but at $10\tau$ the eigenfunctions become severely distorted and the metastable partitions blur.

\subsection{Deca-alanine (Ala\textsubscript{10})}\label{appendix:ala10}

\begin{figure}[h!]
\begin{center}
\includegraphics[width=1.0\textwidth]{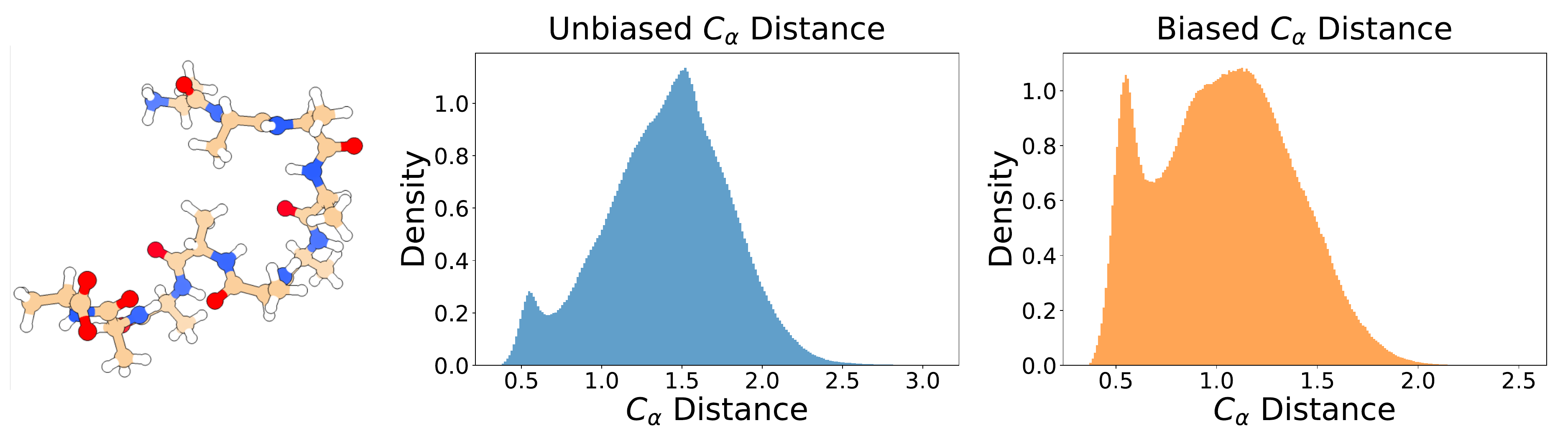}
\end{center}
\vspace{-3mm}
\caption{Distribution of the terminal C\textsubscript{$\alpha$}--C\textsubscript{$\alpha$} 
distance for the Ala\textsubscript{10} system under the unbiased (blue) and biased (orange) 
ensembles. The harmonic umbrella potential centered at $r_0 = 0.5\,\text{nm}$ shifts the 
distribution toward shorter end-to-end distances, enhancing sampling of compact conformations.}
\label{fig:ala10-dist}
\end{figure}

We performed all-atom MD simulations of deca-alanine (Ala\textsubscript{10}), a 103-atom peptide, in implicit solvent. The simulation was carried out with the \textsc{OpenMM}~8.2 simulation package~\cite{eastman2023openmm} at $300\,\text{K}$. The system employed the Amber ff14SB force field~\cite{maier2015ff14sb} with the GB-Neck2 implicit solvent model~\cite{nguyen2013improved} (``\texttt{amber14-all.xml}'', ``\texttt{implicit/gbn2.xml}''). Dynamics were propagated with an underdamped Langevin integrator with a time step of $2\,\text{fs}$. The aggregated simulation time was $5\,\mu\text{s}$ for both the unbiased and biased trajectories. Coordinates and Girsanov reweighting (GR) factors were saved every 200 steps ($400\,\text{fs}$) using the Girsanov-enabled \textsc{OpenMM} implementation~\cite{schafer2024implementation}.

For the biased simulation, a harmonic umbrella potential was applied to the distance between the two terminal C\textsubscript{$\alpha$} atoms:
\begin{equation}
    U(r) = \tfrac{1}{2}\,k\,(r - r_0)^2,
\end{equation}
with spring constant $k = 20\,\text{kJ}\,\text{mol}^{-1}\,\text{nm}^{-2}$ and target distance $r_0 = 0.5\,\text{nm}$. Figure~\ref{fig:ala10-dist} compares the distribution of the C\textsubscript{$\alpha$} end-to-end distance under the unbiased and biased ensembles, confirming that the umbrella potential significantly shifts the sampled distribution toward compact conformations near $r_0$.

To extract slow collective motions and reduce the dimensionality of the feature space, we applied time-lagged independent component analysis (TICA)~\cite{molgedey1994separation, perez2013identification} with a lag time of $0.4\,\text{ns}$, retaining the leading 20 components. These 20-dimensional TICA features were subsequently used both as input for training the MGR reweighting network and as the basis for constructing all Markov state models. The MSM was built by discretizing the TICA space into 500 clusters via $k$-means clustering, and transition matrices were estimated at varying lag times to assess convergence of the implied timescales.

\subsection{Model architecture}
Our classifier estimator $h_{\theta}(\cdot, \cdot)$ in MGR is modeled by 3 layer MLPs, augmented with Fourier feature encodings \cite{tancik2020fourier}. We consider two types of encoding $\gamma(\mathbf v)$, ($\mathbf v=[x_t, x_{t+k\tau}]$ is the concatenation of the paired data):

\textbf{Positional encoding:} $\gamma(\mathbf v) = [\sin(\mathbf v), \cos(\mathbf v), \sin(2\mathbf v), \cos(2\mathbf v), ..., \sin(B\mathbf v), \cos(B\mathbf v)]$, where $B=10$ denotes the scaling number.

\textbf{Gaussian encoding:} $\gamma(\mathbf v) = [\cos(B\mathbf v), \sin(B\mathbf v)]$, where $B\in \mathbb R^{m\times 2d}$ is sampled from $\mathcal{N}(0, 10)$.

Both Fourier-feature variants converge faster during training and yield comparable estimation accuracy. Detailed comparison is reported in Appendix~\ref{appendix:neural-ablation}. In our experiments, we adopt a positional-encoding MLP with ReLU activations.

\subsection{Computational Cost}\label{appendix:cost}
 
We report the computational cost of training and evaluating MGR on the four benchmark systems in Table~\ref{tab:cost}. All experiments were performed on a single NVIDIA GeForce RTX 4090 GPU. The training lag time $\tau$ listed below is the short-lag used during training, and the longer effective lag times are obtained through the iterative extension described in Section~\ref{sec:MGR}. Inference time refers to the wall-clock time required to evaluate $w_{k\tau}$ on all transition pairs used for MSM construction.
 
\begin{table}[h!]
\centering
\footnotesize
\caption{Computational cost of MGR on benchmark systems.}
\label{tab:cost}
\begin{tabular}{lccccccc}
\toprule
System & RC & Traj.\ length & $\tau$ & Epochs & Training & Inference & Model\\
 & Dim. & (steps) & (steps) & per lag & time (h) & time (s) & size (MB) \\
% System & RC & Traj.\ length & $\tau$ & Epochs & Training & Inference & Data gen.\\
%  & Dim. & (steps) & (steps) & per lag & time (h) & time (s) & time (h) \\
\midrule
Four Well          & 1  & $1\times 10^{7}$   & 50   & 20 & 1  & <3 & 2.08 \\
M\"uller--Brown    & 2  & $1\times 10^{7}$   & 10   & 20 & 1  & <3 & 2.16 \\
Alanine dipeptide  & 2  & $2.5\times 10^{8}$ & 20 (40\,fs)  & 20 & 18 & <10 & 2.16 \\
Deca-alanine       & 20 & $1\times 10^{8}$   & 200 (400\,fs) & 20 & 11 & <10 & 3.57 \\
% Four Well          & 1  & $1\times 10^{7}$   & 50   & 20 & 1  & <3 & 2 \\
% M\"uller--Brown    & 2  & $1\times 10^{7}$   & 10   & 20 & 1  & <3 & 3 \\
% Alanine dipeptide  & 2  & $2.5\times 10^{8}$ & 20 (40\,fs)  & 20 & 18 & <10 & 36 \\
% Deca-alanine       & 20 & $1\times 10^{8}$   & 200 (400\,fs) & 20 & 11 & <10 & 40 \\
\bottomrule
\end{tabular}
\end{table}
 
Across all systems, training completes within 18 GPU hours and inference under 10 seconds. These results indicate that MGR adds little overhead compared to the cost of the underlying MD simulations.

\section{Ablation}
\subsection{Neural ratio estimation} \label{appendix:ratio}
Ratio estimation \cite{sugiyama2010density} is a fundamental technique for comparing two distributions. Kernel moment matching, e.g. KMM \cite{gretton2009covariate}, matches all the moments with reproducing kernels, which is effective and computationally efficient. Probabilistic classification recasts ratio estimation as posteriors from a binary classifier \cite{menon2016linking}, showing powerful fitting capability. Featurized classification with normalizing flows \cite{choi2021featurized} further performs classification in a learned latent space, mitigating issues caused by large distributional discrepancies. Path-based methods \cite{choi2022density,yu2025density} connect the two distributions via a continuous probability path and estimate the density ratio by integrating a learned time score. By constructing consecutive path distributions, it alleviates the problems caused by poor overlap between two densities.

However, unlike the standard setting with samples from both distributions, here we only have samples from one distribution plus reference weights linking two distributions. We therefore conducted minor adaptions to estimators below and compare their performance on Four well system. We also adapted the path-based method \cite{yu2025density}, but it exhibited numerical instability in our setting.

\paragraph{Standard classifier (weighted BCE)} Following Section~\ref{sec:ratio}, we train a binary classifier on endpoint pairs $(x_t, x_{t+k\tau})$ with weighted cross-entropy in Eq.~\ref{eq:weightedBCE}.

\paragraph{Featurized classifier (weighted BCE)} 
We first map each paired sample to a latent representation $z_{\phi} = \Phi(x_t, x_{t+k\tau};\phi)$, and then perform the classifier-based ratio estimation in this feature space. 
A joint training objective is adopted \cite{choi2021featurized}:
\[
\mathcal L_{\text{joint}} = \alpha \mathcal L_{\text{BCE}}(\theta, \phi) + (1-\alpha) \mathcal L_{\text{latent}}(\phi),
\]
where $\mathcal L_{\text{BCE}}(\theta, \phi)=- \mathbb E_{t} \left[ c_t \log h_{\theta}(z_{\phi}) + \log \left(1- h_{\theta}(z_{\phi})\right) \right]$ is the weighted binary cross entropy in latent space, $\mathcal L_{\text{latent}}(\phi)$ denotes the objective for optimizing encoder network, and $\alpha=0.5$ is a hyperparameter. 

Here, we consider two encoders to map data into latent space: (i) an invertible normalizing-flow encoder (Classifier-NF) \cite{choi2021featurized}, and (ii) a non-bijective MLP encoder (Classifier-MLP). Classifier–NF guarantees ratios computed in feature space are equivalent to those in input space,
whereas Classifier-MLP uses more flexible, non-invertible networks at the cost of potential information loss. 
For Classifier-NF, we train the encoder by maximum likelihood loss $\mathcal L_{\text{latent}}^{\text{KL}}(\phi)$ \cite{dinh2016density,choi2021featurized}, and for Classifier-MLP, we minimize a sliced 2-wasserstein distance $\mathcal L_{\text{latent}}^{\text{Wass}}(\phi)$\cite{kolouri2018sliced} between the latent variables and a standard Gaussian.

% \paragraph{Path-based method} Since $\rho_{k\tau} \propto c_t\cdot\tilde \rho_{k\tau}$, we define two empirical distributions with given data pairs during training: the first, $p_1$, uniformly samples each paired data from $\tilde \rho_{k\tau}$; the second, $p_2$, samples each pair with normalized weight $c_t$. Then an interpolated path between $p_1$ and $p_2$ is constructed \cite{yu2025density}, and the ratio is estimated by learned the time score \cite{choi2022density}.

\begin{figure}[h!]
\begin{center}
\includegraphics[width=1\textwidth]{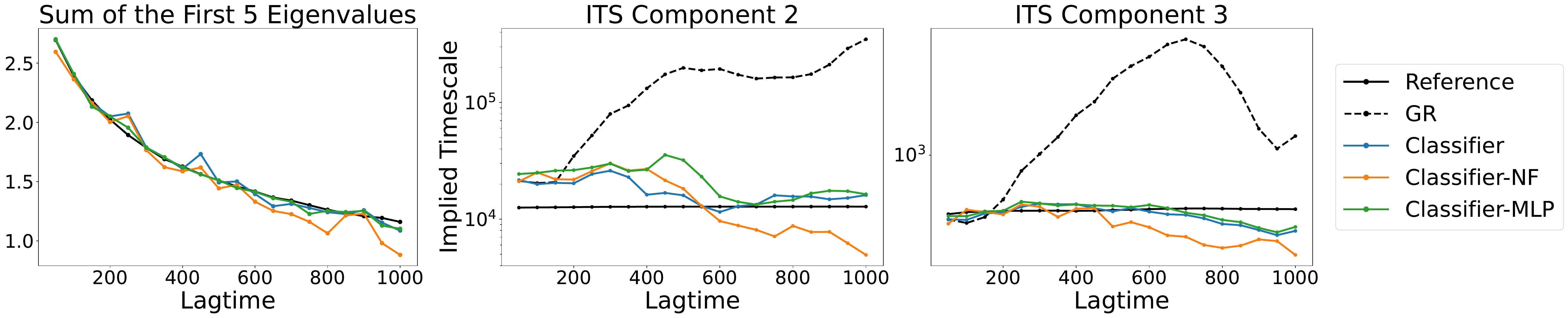}
\end{center}
\vspace{-4mm}
\caption{Neural ratio estimators ablation on the 1D Four well system. \textbf{Left:} Sum of the first 5 eigenvalues as a function of lagtime across different ratio estimators. \textbf{Right:} The first 2 ITS as a function of lagtime across different ratio estimators.}
\label{fig:ratio-ablation}
\end{figure}

Comparative results are reported in Figure~\ref{fig:ratio-ablation}, where featurized classifiers did not show measurable improvement over a standard classifier. We adopt the standard classifier in Section~\ref{sec:ratio}, which provides satisfying results. Alternative advanced ratio estimators, and their applications, require further investigation.

\subsection{Lagtime chosen} \label{appendix:lag-ablation}
Selecting the suitable short lag $\tau$ used in $w_{\tau}^{\text{GR}}$ is crucial for MGR. Large $\tau$ will inflate the variance of Girsanov weights and can destabilize training, while small $\tau$ will require many iterations to reach long timescales, accumulating approximation error and cost. As a practical rule, we recommend choosing $\tau$, which the relative ESS of the Girsanov weights falls in the range $0.15\sim 0.5$. This strikes a balance between weight degeneracy and excessive iteration depth.

\begin{figure}[h!]
\begin{center}
\includegraphics[width=1\textwidth]{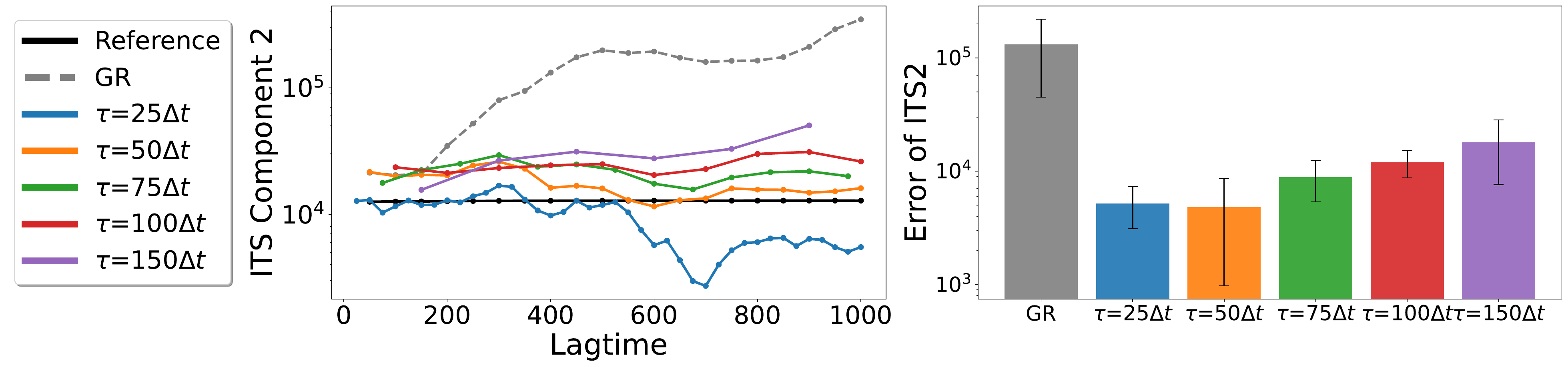}
\end{center}
\vspace{-6mm}
\caption{Lagtime ablation on the 1D Four well system. \textbf{Left:} ITS2 as a function of lagtime for models trained with different short-lag values $\tau$. \textbf{Right:} Mean and standard deviation of the ITS2 error, aggregated over all evaluation lags for models trained with different short-lag values $\tau$.}
\label{fig:lagtime-ablation}
\end{figure}

Take Four well system as an example. We train MGR with short-lag values $\tau \in \{25, 50, 75, 100,$ $ 150\} \Delta t$. The corresponding relative 
$\text{ESS}$ of $w_{\tau}^{\text{GR}}$ is $\{0.73, 0.43, 0.24, 0.11, 0.06\}$. All other settings are kept identical.
We then compare the dominant second implied timescale (ITS2) across evaluation lags in Figure~\ref{fig:lagtime-ablation}. It shows that the model trained under $\tau=50\Delta t$ produces the most stable ITS2 and matched the unbiased (reference) result most closely. As $\tau$ increases, the discrepancy between the model results and the unbiased reference grows.

\subsection{Model architecture} \label{appendix:neural-ablation}
\begin{figure}[h!]
\vspace{-3mm}
\begin{center}
\includegraphics[width=1\textwidth]{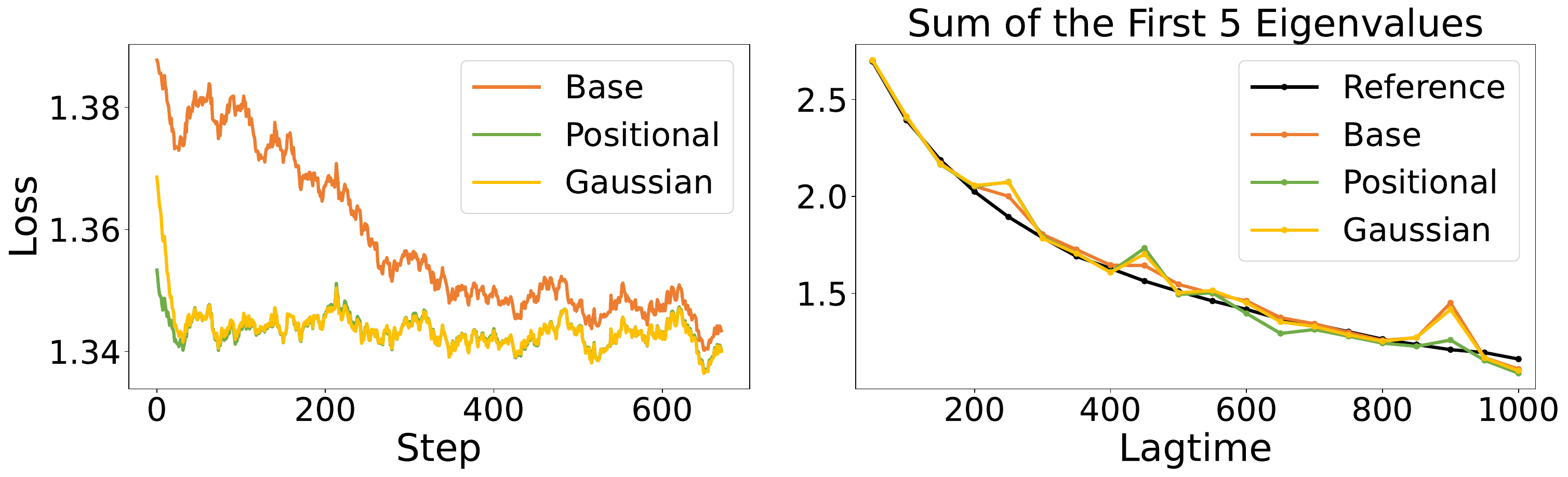}
\end{center}
\vspace{-6mm}
\caption{Network ablation on the 1D Four well system. \textbf{Left:} Training curve as a function of training steps across different networks. \textbf{Right:} Sum of the first 5 eigenvalues as a function of lagtime across different networks.}
\label{fig:network-ablation}
\end{figure}

To demonstrate network efficiency, we compare a plain MLP, a positional encoding MLP, and a Gaussian encoding MLP on Four well system in Figure~\ref{fig:network-ablation}. Both Fourier-feature models (Positional and Gaussian) converge faster than the plain MLP (Base). Three models show comparable estimation accuracy during evaluation.

% \section{The Use of Large Language Models (LLMs)}
% Large Language Models (LLMs) were used solely to polish the writing and grammar of the manuscript. Authors take full responsibility for the paper.

\end{document}

%% file: math_commands.tex
\usepackage{amsmath,amsfonts,bm}

\def\eqref#1{equation~\ref{#1}}
\def\1{\bm{1}}

\DeclareMathAlphabet{\mathsfit}{\encodingdefault}{\sfdefault}{m}{sl}
\SetMathAlphabet{\mathsfit}{bold}{\encodingdefault}{\sfdefault}{bx}{n}